\documentclass[aps,amsmath,amssymb,twocolumn,showpacs,prx,superscriptaddress]{revtex4-1}

\usepackage{graphicx,epsfig}
\usepackage{amsthm}
\usepackage{amsmath}
\usepackage{amssymb}
\usepackage{mathdots}
\usepackage[normalem]{ulem}
\usepackage{natbib}
\usepackage{color}

\renewcommand{\vec}[1]{{\boldsymbol #1}}

\usepackage{color}
\usepackage{ulem}

\begin{document}

\title{Analytical and Numerical study of the out-of-equilibrium current through a helical edge coupled to a magnetic impurity}
\smallskip

\author{Yuval Vinkler-Aviv}
\affiliation{University of Cologne, Institute for Theoretical Physics, 50937 Cologne, Germany}

\author{Daniel May}
\affiliation{Theoretische Physik 2, Technische Universit\"at Dortmund,
  44221 Dortmund, Germany}

\author{Frithjof B. Anders}
\affiliation{Theoretische Physik 2, Technische Universit\"at Dortmund,
  44221 Dortmund, Germany}

\begin{abstract}
We study the conductance of a time-reversal symmetric helical electronic edge coupled antiferromagnetically to a magnetic impurity, employing analytical and numerical approaches. The impurity can reduce the perfect conductance $G_0$ of a noninteracting helical edge by generating a backscattered current. The backscattered steady-state current tends to vanish below the Kondo temperature $T_K$ for time-reversal symmetric setups. We show that the central role in maintaining the perfect conductance is played by a global $U(1)$ symmetry. This symmetry can be broken by an anisotropic exchange coupling of the helical modes to the local impurity. Such anisotropy, in general, dynamically vanishes during the renormalization group (RG) flow to the strong coupling limit at low-temperatures. The role of the anisotropic exchange coupling is further studied using the time-dependent Numerical Renormalization Group (TD-NRG) method, uniquely suitable for calculating out-of-equilibrium observables of strongly correlated setups. We investigate the role of finite bias voltage and temperature in cutting the RG flow before the isotropic strong-coupling fixed point is reached, extract the relevant energy scales and the manner in which the crossover from the weakly interacting regime to the strong-coupling backscattering-free screened regime is manifested. Most notably, we find that at low temperatures the conductance of the backscattering current follows a power-law behavior $G\sim (T/T_K)^2$, which we understand as a strong nonlinear effect due to time-reversal symmetry breaking by the finite-bias.
\end{abstract}

\maketitle

\section{Introduction}
Chiral electronic channels, which can be found on the edges of an integer quantum Hall sample, show unique conductance behavior. As backscattering of electrons is not possible, the conductance of these channels is robust against many perturbations, {\it inter alia} scattering off impurities, and it attains the universal value of $G_0=e^2/h$ per charge-carrying channel. While a system with a single chirality requires breaking of time-reversal symmetry, as in the quantum Hall effect, a more nuanced picture emerges when one considers helical modes. In these systems, the spin and propagation direction are interlinked, with opposite flavors of spins counter-propagating. For example, the edges of a topological insulator such as a quantum spin-Hall bar demonstrate this behavior, without breaking time-reversal symmetry~\cite{Bernevig_2006,Kane_2005,Qi_2011,Hasan_2010}. Such systems have focused a great amount of interest in recent years, both experimentally and theoretically. One of the signatures of the quantum spin-Hall state should be a perfect edge conductance at low temperatures and bias voltages when time-reversal symmetry is maintained, as backscattering of electrons along the edge requires flipping of the spin, which is strongly suppressed in presence of time-reversal symmetry.

Experimentally, however, the perfect quantization of the conductance was not observed, despite measurements in different topological insulators such as HgTe/CdTe and InAs/GaSb quantum wells, bismuth layers and WTe$_2$ monolayers~\cite{Jia_2017,Fei_2017,Sabater_2013,Mueller_2015,Li_2017,Suzuki_2015, Du_2015,Spanton_2014,Suzuki_2013, Olshanetsky_2015,Knez_2011,Gusev_2013,Gusev_2014a,Gusev_2014b, Grabecki_2013,Nowack_2013,Kononov_2015,Brune_2012,Roth_2009,Konig_2007}. Suggestions for the potential sources for the deviation from perfect conductance include effects such as electron-electron interactions, disorder, electrical noise, inelastic scattering, and others~\cite{Xu_2006, Stroem_2010, Schmidt_2012, Kainaris_2014, Vayrynen_2018, Vayrynen_2013, Rod_2015, Aseev_2016, Wang_2017, Hsu_2017, Hsu_2018}.

The question of the effect of magnetic impurities on the conductance along helical edges was the subject of theoretical attention as well, considering different forms of impurities, coupling, and electronic band structures~\cite{Lezmy_2012, Eriksson_2012, Altshuler_2013, Hattori_2014, Yevtushenko_2015, Vayrynen_2016,  Yevtushenko_2018, Vayrynen_2014, Maciejko_2009, Maciejko_2012, Chelanov_2013, Kimme_2016, Kurilovich_2017a, Kurilovich_2017b, Kurilovich_2019}. At low temperatures and in the absence of strong electron-electron interactions, a generic magnetic impurity forms a Kondo singlet and is screened out, allowing the helical edge to reconstitute itself around it and, therefore, has no effect on the conductance. This has been the fundamental picture established by Wu and collaborators and by Maciejko and collaborators~\cite{Wu_2006,Maciejko_2009}. However, identifying the leading corrections at finite temperatures to the perfect conductance is an ongoing subject for debate.

In Ref.~\cite{Maciejko_2009}, the authors employed bosonization and analytical perturbative RG calculations in order to study the backscattering from a magnetic impurity, and predicted that at low temperatures the deviation from perfect conductance scales as $G \propto (T/T_K)^{2(4K-1)}$ as long as $K>1/4$, where $K$ is the Luttinger parameter describing the strength of the electron-electron interactions along the edge, and $T_K$ the Kondo temperature. Specifically, for noninteracting electrons ($K=1$), $G \propto (T/T_K)^6$ is found. V\"ayrynen and collaborators~\cite{Vayrynen_2013} studied the conductance in presence of charge puddles created by disorder and modeled by a series of interacting quantum dots. They reported a deviation from perfect conductance in the linear bias voltage regime and for low temperatures due to a backscattering current with a condutance behavior of $G \propto T^4$. Recently, Kurilovich and collaborators considered coupling to an impurity spin with $S>1/2$, and focused on the effect of the local spin anisotropy on the conductance~\cite{Kurilovich_2017a,Kurilovich_2019}. They discovered that this effect is strongly dependent on whether the spins is integer of half-integer, and that the correction is almost temperature independent down to low temperatures.

As Tanaka and collaborators~\cite{Tanaka_2011} argued, the isotropic Kondo coupling alone does not affect the perfect dc conductance for any $K$ and temperature $T$. They showed that this can be understood due to the fact that time-reversal symmetry allows backscattering only accompanied with a spin-flip of the impurity, which can be further flipped back only with backscattering in the opposite direction, thus prohibiting a steady-state backscattered current. In order to circumvent this limitation while preserving time-reversal symmetry, one has to consider an anisotropic exchange coupling~\cite{Kurilovich_2017a,Kurilovich_2017b, Kurilovich_2019,Vayrynen_2016} or describe coupling to a many-level interacting quantum dot~\cite{Chelanov_2013, Vayrynen_2013, Vayrynen_2014}.

While a plethora of theoretical tools was employed to study the effects of magnetic impurities on the conductance in helical systems, to the best of our knowledge the problem was not yet addressed using advanced numerical tools, despite the large success of such methods, {\it e.g.} the numerical renormalization group (NRG), in exploring the features of strongly-correlated impurity models~\cite{Bulla2008}. In this paper we employ the NRG and time-dependent NRG (TD-NRG) technique to study the conductance of a helical edge coupled to an impurity in non-equilibrium steady state, when finite bias voltage is applied, over a range of temperatures and exchange couplings.

The structure of the paper is as follows. We start in Sec.~\ref{sec:model_and_current} by presenting the model Hamiltonian, deriving the expressions for the current in terms of the non-equilibrium Green's functions of a local degree of freedom and analyze its character. In Sec.~\ref{sec:perturbative_RG} we employ perturbative RG methods to analytically study the structure of the correlations and how they affect the conductance. Then, in Sec.~\ref{sec:td_NRG}, we turn to the advanced numerical method of TD-NRG to calculate the current through the helical modes for different temperatures, bias voltages, and interactions. Finally, in Sec.~\ref{sec:discussion} we discuss our results and their implications.

\section{Model and Observables}
\label{sec:model_and_current}
\subsection{The Hamiltonian and its symmetries}
We consider the 1d edge of a quantum spin-Hall insulator, which is characterized by two counter-propagating helical electronic modes, associated with two opposite spin projections and described by the field operators $\psi_{\sigma}(x)$. The edge electrons are coupled at the origin $x=0$ to a set of local fermionic degrees of freedom $D_{n,\sigma}$ which describes a local interacting impurity. For the time being, we will not consider specific interaction terms, and discuss the setup in general. The only requirement we shall impose is that the entire setup is time-reversal symmetric, which is satisfied by the helical modes as long as $\psi_{\sigma}$ and $\psi_{-\sigma}$ are a Kramers pair, and that they accordingly couple to Kramers pairs degrees of freedom of the impurity.

In reality, the helicity in the edge of quantum spin-Hall insulator comes from spin-orbit coupling, which means that although the left- and right-moving electrons have opposite spin projections at each point, that spin projection is not constant along the edge. This was suggested as a possible backscattering mechanism, allowing for momenta-dependent flipping of the spin through inelastic scattering processes or the Dyakonov-Perel spin relaxation mechanism~\cite{Schmidt_2012,Pala_2005}. As we are interested in the effects of the impurity on the conductance, we neglect this effect and assume that the spin orientation is constant along the edge. This can be formally achieved by applying a space dependent unitary transformation that rotates the spins at each point to the same direction, and then omitting the extra momenta-dependent terms that result from this transformation.

The Hamiltonian that describes the dynamics of the edge electrons is given by
\begin{equation}
	\mathcal{H}_{\rm e} = -iv_F\sum_{\sigma}
					\sigma \int\! dx \psi^{\dagger}_{\sigma}(x)
					\partial_x \psi_{\sigma}(x),
\end{equation}
with $\sigma = +1$ ($\sigma=-1$) for right (left) movers, which also have opposite spins. For convenience, and without loss of generality, we shall henceforth identify the right-movers with up-spins in the $z$-directions and left-movers with down-spins. Under time-reversal transformation $\hat{T}$, the fields undergo $\hat{T}\psi_{\sigma}\hat{T}^{-1} = \sigma\psi_{-\sigma}$.

At the origin $x=0$, the edge electrons hybridize with the degrees of freedom of a local impurity $D_{n,\sigma}$, which might have more than one level (orbital) per spin
\begin{equation}
\mathcal{H}_t = \sum_{\sigma,n} t_{\sigma,n}\psi^{\dagger}_\sigma(0)D_{n,\sigma} + {\rm h.c.},
\end{equation}
where $n=1,\ldots,N$ labels the impurity levels, and $t_{n,\sigma}$ the hybridization parameters. The levels of the impurity are also arranged in time-reversal symmetric pairs $\hat{T}D_{n,\sigma}\hat{T}^{-1} = \sigma D_{n,-\sigma}$, where the time-reversal symmetry enforces $t_{n,\sigma} = t^*_{n,-\sigma}$. It is convenient to define a single degree of freedom $d_{\sigma}$ with which each spin-flavor of the edge electrons hybridize 
\begin{eqnarray}
d_{\sigma} &=& t_{\sigma}^{-1}\sum_{n} t_{n,\sigma} D_{n,\sigma}
\end{eqnarray}
with $t_{\sigma} = \sqrt{\sum_n |t_{n,\sigma}|^2}$, and construct an orthogonal set describing all the other $N-1$ levels $D^{o}_{n,\sigma}$. Then
\begin{equation}
	\mathcal{H}_t = \sum_{\sigma} t_{\sigma} 
	\psi^{\dagger}_{\sigma}(0) d_{\sigma} + {\rm h.c.}.
\end{equation}
In the general case where the $D$-levels are non-degenerate, this transformation leads to extra terms between the impurity levels themselves.

\begin{figure}
\includegraphics[scale=0.6]{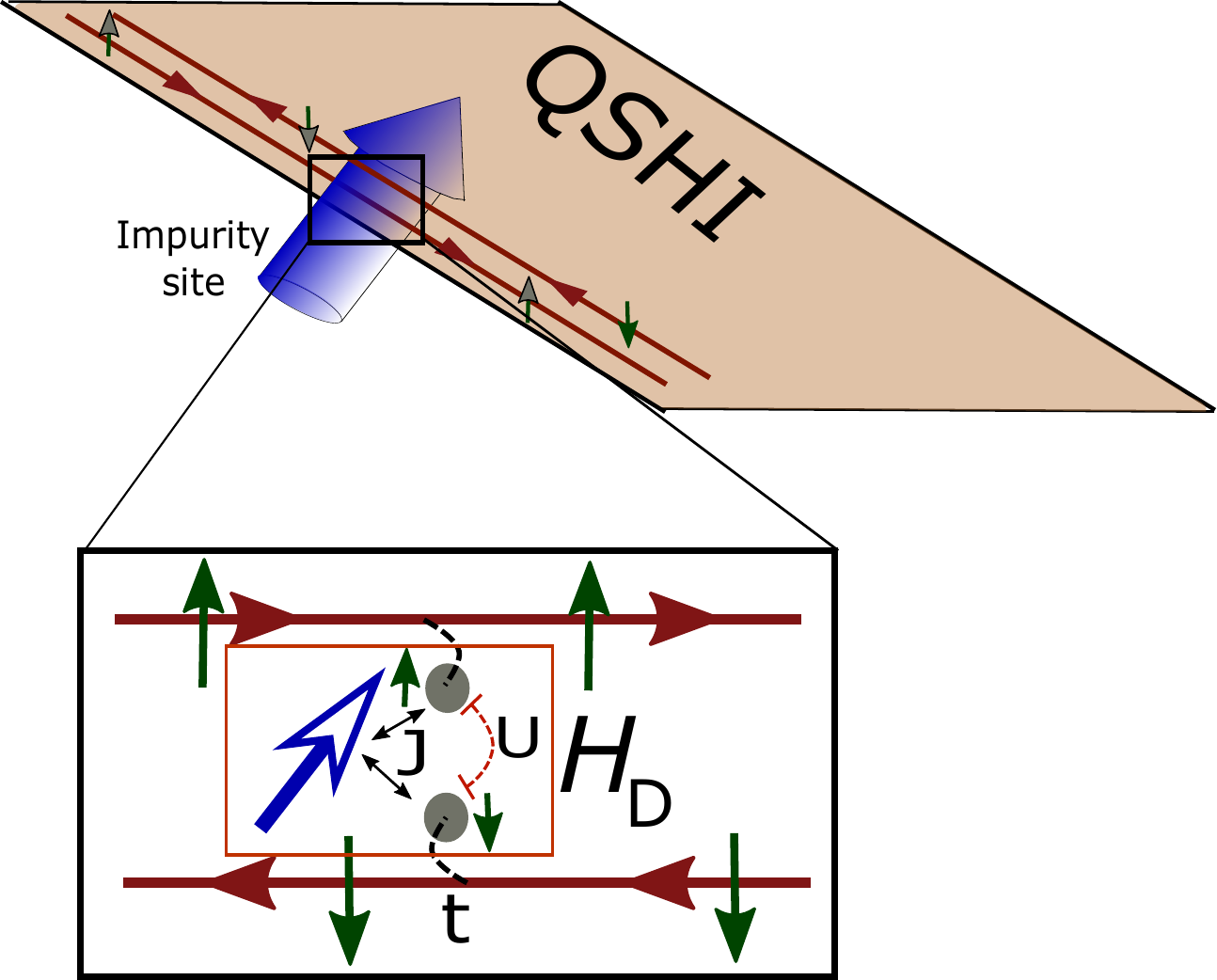}
\caption{Schematic depiction of the setup considered throughout most of this paper. A 1d helical edge of a quantum spin Hall insulator consists of right-moving up-spin electrons and left-moving down-spin electrons, coupled at a single point to an generalized impurity via tunnel amplitude $t$. This impurity encompasses a correlated spinful level interacting with an additional $S=1/2$ quantum spin. Time reversal symmetry is maintained by making the up and down levels Kramers partners, and keeping a real exchange coupling elements $J$ with the impurity spin.}
\label{fig:setup}
\end{figure}

The dynamics of the impurity degrees of freedom, and of potentially other local degrees of freedom that interact with the $D_{n,\sigma}$ orbitals, are described by a general interacting Hamiltonian $H_{D}$, which does not contain $\psi_{\sigma}$, and does not violate time-reversal symmetry. The full Hamiltonian is $\mathcal{H} = \mathcal{H}_{\rm e}+\mathcal{H}_t + \mathcal{H}_D$, and by construction it is time-reversal symmetric. A schematic depiction of the setup is given in Fig.~\ref{fig:setup}, where we assumed energy degenerate impurity orbitals.

\subsection{Electric current}
\label{sec:current}
In this section we derive the relevant Meir-Wingreen expression~\cite{Meir_1992} for the electrical current through the edge in terms of the local Green's functions of the localized level $d_{\sigma}$. We analyze its properties and compare it with the expression for the current through a non-helical 1d system.

In absence of a coupling to the localized level, $t_{\sigma}=0$, the number of right-moving electrons $\hat{N}_{R}$ and left-moving electrons $\hat{N}_{L}$ is constant, and the steady-state current is given by the difference in the corresponding densities $\hat{I}_0 = v_F e \left(\hat{n}_{R}-\hat{n}_{L}\right)$ with $\hat{n}_{R/L}$ the densities of the left and right movers. Plugging in the density of states per unit length $\rho_0 = 1/(2\pi v_F)$ and integrating over the different occupancies we arrive at the standard result
\begin{align}
  \begin{split}
    \label{eqn:5}
	I_0 &= \langle \hat{I}_0 \rangle = e \int\! \frac{d\epsilon}{2\pi} \left[ f(\epsilon-\mu_+)-f(\epsilon-\mu_-) \right]\\
    &\simeq  G_0\frac{\mu_+-\mu_-}{e}
  \end{split}
\end{align}
with $\mu_{\pm}$ the chemical potential of the left and right movers, $f(\epsilon)$ the Fermi-Dirac distribution, and we assumed a large electronic bandwidth $D\gg |\mu_{\pm}|,T$. The perfect conduction of the clean channel may be reduced by a backscattered current $\hat{I}_{B}$ that takes a right moving particle and reflects it into a left moving one $\hat{I} = \hat{I}_0-\hat{I}_B$. The symmetric form of the backscattered current operator is given by
\begin{align}
  \begin{split}
    \label{eq:I-B}
	\hat{I}_{B} &= \frac{e}{2}\frac{d}{dt}\left(\hat{N}_{L}-\hat{N}_{R}\right) \\
	&= i\frac{e}{2}\left[t_{-}\psi^{\dagger}_-(0)d_{-}- t_{+}\psi^{\dagger}_+(0)d_{+} - {\rm h.c.}\right].
  \end{split}
\end{align}
In order to evaluate $I_B = \langle \hat{I}_B\rangle$ at steady-state, we express it using the lesser Green's functions $G^<_{AB}(\tau,\tau') = \langle B(\tau ')A(\tau) \rangle$ which are functions only of the time difference $\tau-\tau '$ at steady-state. Upon Fourier transforming with respect to the time difference we arrive at
\begin{equation}
	I_B = e \; \Im\int\! \frac{d\omega}{2\pi}
	\left[t_{+}G^<_{d_{+}\psi^{\dagger}_{+}}(\omega)-
	t_{-}G^<_{d_{-}\psi^{\dagger}_{-}}(\omega)\right],
\end{equation}
and by applying standard diagrammatic expansion we obtain $$G^<_{d_{\sigma}\psi^{\dagger}_{\sigma}}(\omega) = t^*_{\sigma}[G_{d_{\sigma}d^{\dagger}_{\sigma}}(\omega)g_{\psi_{\sigma}\psi^{\dagger}_{\sigma}}(\omega)]^<.$$ Here $g(\omega)$ is the bare Green's function taken with respect to $\mathcal{H}_{\rm e}$, whereas $G(\omega)$ signifies the Keldysh Green's function in presence of the full Hamiltonian $\mathcal{H}$. We are to take the lesser part of the product of the two Green's functions, which is realized by applying Langreth's rules~\cite{Langreth_76}. The bare Green's functions of the electrons at the edge in the wide-band limit are given by
\begin{align}
  \begin{split}
	g^{r/a}_{\psi_{\sigma}\psi^{\dagger}_{\sigma}}(\omega) &= \mp i\pi\rho_0,\\
	g^{<}_{\psi_{\sigma}\psi^{\dagger}_{\sigma}}(\omega) &=	2\pi\rho_0 f(\omega-\mu_{\sigma}),
  \end{split}
\end{align}
with $g^{r(a)}$ the retarded (advanced) bare Green's function. We similarly label the fully dressed retarded (advanced) Green's function by $G^{r(a)}$. Using these functions and labels we express the backscattered current using only the fully dressed Green's functions of the $d_{\sigma}$ orbitals
\begin{align}
\begin{split}
  I_B =& \frac{G_0}{e}\Gamma \int\! d\omega \bigg[ G^<_{d_{+}}(\omega) + 2{\Im}\{ G^r_{d_{+}}(\omega)
  \}f(\omega-\mu_{+}) \\
  & - G^<_{d_{-}}(\omega) - 2{\Im}\{ G^r_{d_{-}}(\omega) \}f(\omega-\mu_{-}) \bigg].
	\label{eq:I_b_from_G}
\end{split}
\end{align}
Here, $\Gamma = \pi\rho_0 |t_{\sigma}|^2$ equals half the tunneling rate to the localized impurity orbital, which is identical for both $\sigma$ due to time-reversal symmetry, and we used the short-hand notation $G^\nu_{d_{\sigma}}$ for the Green's functions $G^\nu_{d_{\sigma}d^{\dagger}_{\sigma}}$.

Eq.~\eqref{eq:I_b_from_G} is a central result of this section, as it is an exact expression for the non-equilibrium current through the edge $I=G_0 V - I_B$, driven by an applied voltage drop $eV=\mu_+-\mu_-$. It can be evaluated by calculating the fully dressed Green's functions of the localized orbitals alone. No approximations were needed in its derivation from our Hamiltonian, and it encodes all the information about the correlations and temperature dependence through the structure of the fully dressed Green's functions. Note, that the vanishing of the backscattered current is equivalent to $\langle \psi^{\dagger}_{\sigma}(0)d_{\sigma}\rangle = \langle d^{\dagger}_{\sigma} \psi_\sigma(0)\rangle $ implying that these expectation values are real. We now turn to a qualitative discussion, and point out the unique features of the helical edge.

The total current $I$ is a current through a 1d mode which is {\em side-coupled} to an interacting region. Studies of transport in 1d channels side-coupled to an impurity in the Kondo regime have shown that such impurities suppress the conduction completely at low temperatures, in contrast to the perfect transmission when tunneling {\em through} Kondo correlated impurity~\cite{Kang_2001}. However, the setups considered for these studies were markedly different than the setup described here, as both left and right movers carried both spin flavors, and respectively coupled to the Kondo impurity. In the helical edge setup, on the other hand, left and right movers correspond to different spin flavors. To illustrate the difference between these setups, which directly affects the current, we note that the helical edge Hamiltonian cannot be derived from a corresponding 1d lattice model when taking the continuum limit, and it is fundamentally different than the non-helical case. One has to bear in mind that the full model of the quantum spin-Hall insulator is 2d and the helical edge states are effective 1d topologically protected transport channels that can be spatially deformed. Therefore, the strong-coupling picture where side-coupling to an impurity cuts a 1d wire into two pieces, as the site near the impurity hybridizes strongly with it, is not applicable.

On the other hand, the backscattered current $I_B$ describes a current contribution from source to drain {\it through} the impurity, and can be mapped onto a spinless model where two noninteracting leads are coupled through an interacting region. In this mapping the up-spin electrons in the edge are mapped onto a source lead, while the down-spin electrons are the drain. The requirement of time-reversal symmetry in the original Hamiltonian greatly restricts the type of terms allowed in the interacting region. Specifically, levels coupled to the source $d_{+}$ and levels coupled to the drain $d_{-}$ cannot directly be linked as the term $\lambda d^{\dagger}_{+}d_{-}$ breaks time-reversal symmetry. In order to get non-vanishing backscattered current in steady state one must overcome this obstacle by considering additional interaction terms.

\subsection{$U(1)$ symmetry and the current}
\label{sec:symmetry}
In this section we define a $U(1)$ symmetry the system might maintain, and demonstrate its importance in protecting the perfect conductance of the edge even for finite bias and temperatures. We show that without explicitly breaking this symmetry no steady-state backscattered current can be driven by the local impurity. This is demonstrated by applying a time-dependent gauge transformations, and separately by employing Hershfield's $Y$-operator formalism.

While the $SU(2)$ symmetry is broken by the helical states, we can define a global $U(1)$ symmetry in absence of $\mathcal{H}_D$. The transformation $\psi_{\sigma} \to e^{i\sigma\theta}\psi_{\sigma}$, $d_{\sigma} \to e^{i\sigma\theta}d_{\sigma}$ leaves both $\mathcal{H}_{\rm e}$ and $\mathcal{H}_t$ invariant and preserves time-reversal symmetry. This symmetry is equivalent to a global rotation about the joined spin $z$-axis of the electrons at the edge and the $d_{\sigma}$ orbital. This can be further generalized to encompass degrees of freedom included only in $\mathcal{H}_D$. By summation, one can construct $\mathcal{S}^z = S^z_{\rm mac}+S^z_{\rm mic}$ with 
\begin{align}
  \begin{split}
    \label{eq:U1_generator}
	S^z_{\rm mac} &= \sum_{\sigma}\sigma \int\!dx \psi^{\dagger}_{\sigma}(x)\psi_{\sigma}(x),\\
	S^z_{\rm mic} &= \sum_{\sigma}\sigma \left[d^{\dagger}_{\sigma}d_{\sigma} + \sum_{n} D^{\dagger\;o}_{n,\sigma}D^{o}_{n,\sigma}\right] +2\sum_j S^z_j,
  \end{split}
\end{align}
where ${\bf S}_j$ are the different possible spins degrees of freedom describing the impurity. Then the $U(1)$ rotation is generated by $\exp[i\theta \mathcal{S}^z/2]$. We have either $[\mathcal{S}^z,\mathcal{H}]=0$ for the $U(1)$ symmetric case, or $[\mathcal{S}^z,\mathcal{H}] \neq 0$ when it is broken by $\mathcal{H}_D$.

We begin by applying a gauge transformation using the $U(1)$ generator of Eq.~(\ref{eq:U1_generator}) $U_z(\tau) = \exp[-i\mathcal{S}^z(\mu_+-\mu_-)\tau/2]$, transforming each of the operators according to their charge under $\mathcal{S}^z$. Following the transformation, an extra term is added to the Hamiltonian, given by
$$\Delta \mathcal{H} = i\left[\partial_\tau U_z(\tau)\right]U^{\dagger}_z(\tau),$$ which has a double effect. It shifts the energies of the left- and right-moving edge electrons and eliminates the chemical potential, and in addition, a local effective magnetic field is generated
\begin{equation}
  \mathcal{H}_{B_{\rm eff}} = \frac{\mu_+-\mu_-}{2} S^z_{\rm mic}.
\end{equation}
Operators and expectation values may acquire an explicit time-dependence, which reflects the fact that the setup is out of equilibrium.

In case the $U(1)$ symmetry is maintained, the Hamiltonian and the current operator remain time-independent after the transformation. Since the Hamiltonian and the current operator are time-independent, the problem is mapped onto an effective equilibrium problem, in presence of the local magnetic field, and all expectation values can be calculated with respect to the transformed Hamiltonian. In equilibrium, the fluctuation dissipation theorem ensures that $G^<(\omega) = -2\Im\{G^r(\omega)\} f(\omega)$ which renders the backscattered current in Eq.~(\ref{eq:I_b_from_G}) identically zero at steady-state.

As $\mathcal{S}^z$ is a conserved quantity in this case, and each backscattering event changes the values of $S^z_{\rm mac}$ by $\pm 2$, the values of the local  $S^z_{\rm mic}$ must change accordingly by $\mp 2$ with each backscattering event. Therefore, the coupling of the local degrees of freedom to the effective magnetic field ensures that each backscattering event costs or gains the correct amount of energy $\mu_{\sigma}-\mu_{-\sigma} = eV$. One can also use this fact to convince oneself that the backscattered current must be zero at steady-state: Since $S^z_{\rm mic}$ is a local {\it microscopic} quantity, as long as $\mathcal{S}^z$ is a conserved quantity, $S^z_{\rm mic}$ can allow only a finite number of consecutive backscattering events in the same direction before reaching its maximal allowed value, blocking any further backscattering in that direction.

The situation is starkly different if the $U(1)$ symmetry is broken. In that case, while the current operator following the transformation is still time-independent, the Hamiltonian is bound to be explicitly dependent on time. The setup cannot be described any longer by an effective equilibrium Hamiltonian, and $I_B$ may attain a non-zero value.

A different proof (but similar in spirit) can be constructed by employing the $Y$ operator formalism developed by Hershfield~\cite{Hershfield_1993} to describe non-equilibrium steady-state. In this formalism, the system is described in the distant past $t \to -\infty$ by the density matrix 
\begin{eqnarray}
\rho_0 &=& \frac{1}{Z_0}e^{-\beta(\mathcal{H}_0-Y_0)}
\end{eqnarray}
with $Y_0$ the non-equilibrium condition, and then an interaction term $\mathcal{H}_I$ is turned on adiabatically. The system evolves in time until steady-state is reached. The steady-state density matrix is given by a similar form,
\begin{eqnarray}
\rho &=& \frac{1}{Z}e^{-\beta(\mathcal{H}-Y)}
\end{eqnarray}
with $Y=\sum_{n=0}^{\infty} Y_n$, where $Y_n$ maintains 
\begin{eqnarray}
[\mathcal{H}_0,Y_n]-i\eta Y_n = [Y_{n-1},\mathcal{H}_I]
\end{eqnarray}
for infinitesimal $\eta\to 0^+$.

Hershfield \cite{Hershfield_1993} decomposed the $Y$-operator into the general many-body scattering states operators $\Gamma_{k\sigma}$:
\begin{eqnarray}
Y = \sum_{k\sigma} \mu_\sigma \Gamma^\dagger _{k\sigma}\Gamma _{k\sigma}
\end{eqnarray}
where $\Gamma^\dagger _{k\sigma}$ is expanded in contributions $\Gamma^\dagger _{k\sigma,n}$ proportional to 
the interaction term $(\mathcal{H}_I)^n$ of the Hamiltonian
\begin{eqnarray}
 \Gamma^\dagger _{k\sigma} &=& \sum_{n=0}^\infty  \Gamma^\dagger _{k\sigma,n}
\end{eqnarray}
and each component $\Gamma^\dagger _{k\sigma,n} (n>0)$ obeys the hierarchical differential equation
\begin{eqnarray}
\label{eq:gamma-DGL-17}
\frac{d  \Gamma^\dagger _{k\sigma,n}(t)}{dt} -i \epsilon_{k\sigma}\Gamma^\dagger _{k\sigma,n} &=&
i [ \Gamma^\dagger _{k\sigma,n-1}, \mathcal{H}_I ].
\end{eqnarray}
In order to shed some light into the nature of the $Y$-operator for the $U(1)$ symmetric case, we can evaluate the commutators  in lowest order.
Let us start with $\mathcal{H}_D=0$, $\mathcal{H}_0=\mathcal{H}_e$ and treat the bilinear term $\mathcal{H}_t$ as interaction.
The equations can be analytically solved  yielding the single-particle Lippmann-Schwinger states operators stated below in Eq.\ \eqref{eq:21-LP-single-particle}  ($\Gamma^\dagger _{k\sigma}= \gamma^{\dagger}_{k,\sigma}$). The term $\Gamma^\dagger _{k\sigma}\Gamma _{k\sigma}$ counts the number of fermions in the system with a spin $\sigma$ projection, hence
\begin{eqnarray}
\label{eq:Y-prop-Sz}
Y &=& \frac{\mu_+-\mu_-}{2}\mathcal{S}^z- \frac{\mu_+ +\mu_-}{2}\hat{N}_{\rm tot}
\end{eqnarray}
where $\hat{N}_{\rm tot}$ counts the total number of fermions in the system, and the scattering states operators $\gamma^{\dagger}_{k,\sigma}$ can be used
to write the Hamiltonian $\mathcal{H}_e +\mathcal{H}_t$ in energy diagonal form.

Now we add a finite $\mathcal{H}_D$ that is conserving the total spin component $\mathcal{S}^z$, typically an anisotropic Heisenberg term. The number of left- and right-movers are no longer individually conserved, 
and these states mix due to the interaction in Eq.\ \eqref{eq:gamma-DGL-17}. However, each  mixing term is always associated with a local spin-flip operator 
$S^\pm$, so that the contribution $\Gamma^\dagger _{k\sigma,n}$ maintains its spin excitation character in all orders of the hierarchy 
so that $ \Gamma^\dagger _{k\sigma}$ remains an eigenoperator of the total spin component $\mathcal{S}^z$. Now $\Gamma^\dagger _{k\sigma}\Gamma _{k\sigma}$ counts the number
of spin $\sigma$ excitations in the system and Eq.\ \eqref{eq:Y-prop-Sz} remains valid even for $\mathcal{H}_D\not =0$ as long as $[\mathcal{H},\mathcal{S}^z]=0$.
Note, that one can either construct $\gamma^{\dagger}_{k,\sigma}$ for $\mathcal{H}_D=0$ and then perform a second step by setting
 $\Gamma^\dagger _{k\sigma,0}=\gamma^{\dagger}_{k,\sigma}$ and switch on  $\mathcal{H}_D=0$, or one starts directly from free edge states
 and use $\mathcal{H}_I= \mathcal{H}_t +\mathcal{H}_D$ to arrive at the same final $\Gamma^\dagger _{k\sigma}$.

The density operator is equivalent to the equilibrium operator in a finite magnetic field since the first term in $Y$ corresponds to a global magnetic field applied in $z$-direction. The second term control the overall filling with fermions and can be essentially dropped. Note that while the occupation numbers are governed by $\rho \propto \exp[-\beta(\mathcal{H}- Y)]$, the dynamics is only controlled by the Hamiltonian $\mathcal{H}$ itself. This is important for calculating the Green's functions. One can either carry out an equilibrium calculation with respect to $\mathcal{H'}=\mathcal{H}- Y$ and perform a frequency shift by $\mu_\sigma$
by hand at the end, or use the definition of the Heisenberg operator $O(t) = \exp[i \mathcal{H} t] O \exp[-i \mathcal{H}t]$ to obtain the correct frequency spectrum. We adopted the later scheme since it remains valid in true non-equilibrium situations when the $U(1)$ symmetry is broken.

In the pseudo-equilibrium situation where the $U(1)$ symmetry holds, the spectral functions obey the dissipation-fluctuation theorem and, therefore, the backscattering current $I_B$ vanishes identically.
Although the operators  $\Gamma^\dagger _{k\sigma}$ contain mixing of left- and right-movers, the mixing cannot induct a steady-state backscattering current.
This can be understood in a consecutive application of $\Gamma^\dagger _{k\sigma}$ onto some arbitrary many-body quantum state. Since each
backscattering term is associated with a local spin-flip term, and the local spin has a finite length, these backscattering terms do not contribute in higher order since they lead to a  nil state or to an equal number of back and forth scattering such that the net current always vanished.
This is fundamentally different of a $U(1)$ symmetry breaking interaction.

In conclusion, we showed breaking the $U(1)$ symmetry defined by $\mathcal{S}^z$ of Eq.~(\ref{eq:U1_generator}) is critical in order for the local impurity to drive a backscattering current at steady-state. When $[\mathcal{S}^z,\mathcal{H}]=0$, the system can always be mapped onto an effective equilibrium setup, which leads to a vanishing backscattering current [given in Eq.~(\ref{eq:I_b_from_G})] due to the fluctuation-dissipation theorem. Following this mapping, the non-equilibrium condition plays a role of a magnetic field. Therefore, we must introduce into $\mathcal{H}_D$ terms that do not commute with $\mathcal{S}^z$ in order to obtain finite backscattering current.

\section{Interaction Hamiltonian and perturbative RG analysis}
\label{sec:perturbative_RG}
From now on forward we shall consider a specific form of interaction for $\mathcal{H}_D$. If one considers a localized impurity spin-$1/2$ which interacts with a single spinfull $d$-level, then the most general interaction Hamiltonian that respects time-reversal symmetry is given by
\begin{align}
  \begin{split}
	\mathcal{H}_D =& \epsilon_d \sum_{\sigma=\pm} \hat{n}_\sigma + U\hat{n}_+\hat{n}_-\\
    &+ \sum_{\alpha,\beta,\nu,\nu'}J_{\alpha,\beta}S^{\alpha}d^{\dagger}_{\nu} \sigma^{\beta}_{\nu,\nu '}d_{\nu'}.
  \end{split}
\end{align}
Here, $\hat{n}_\sigma = d^{\dagger}_\sigma d_\sigma$, $\sigma^\beta_{\nu,\nu'}$
are matrix elements of the Pauli matrices and $J_{\alpha,\beta}$ is a set of nine real coupling coefficients. We used the indices $\nu,\nu'$ in this sum for the helical label $\sigma$ in order to distinguish the label from the symbol for the Pauli matrices. The first two terms describe the on-site energy and Coulomb repulsion between the levels, while the last term is a time-reversal symmetric exchange coupling between the spinfull $d$-level and the impurity spin. We note that when considering the case of an impurity spin with spin larger than $1/2$, the Hamiltonian may also include spin-anisotropy terms $M_{\alpha}(S^{\alpha})^2$ which are nontrivial. These terms may play an important role in driving backscattering current in such setups~\cite{Kurilovich_2017a,Kurilovich_2019}.

\subsection{Mapping onto the anisotropic Kondo Hamiltonian}
It is instructive to map the Hamiltonian onto the well-studied Kondo Hamiltonian. To this end, we start by diagonalizing $\mathcal{H}_e+\mathcal{H}_t$ exactly using the helical scattering states, given by
\begin{align}
  \begin{split}
    \label{eq:21-LP-single-particle}
	\gamma^{\dagger}_{k,\sigma} =& e^{i\phi_k}\psi^{\dagger}_{k,\sigma} +	\frac{t_{\sigma}}{\sqrt{2\pi}} \vert g_{d_{\sigma}}(\epsilon_k+i\eta)\vert\\
    &\times\left(d^{\dagger}_{\sigma} + \int\! \frac{dk'}{\sqrt{2\pi}} \frac{t_{\sigma}}	{\epsilon_k-\epsilon_{k'}+i\eta}\psi^{\dagger}_{k',\sigma}\right),
  \end{split}
\end{align}
that can be derived from Eq.~\eqref{eq:gamma-DGL-17}~\cite{Lebanon_2003}.
Here, $$g_{d_{\sigma}}(z) = \left[z-\int\! \frac{dk}{2\pi} \frac{|t_\sigma|^2}{z-\epsilon_k}\right]^{-1}$$ is the Green's function associated with the level $d_\sigma$, and $\phi_k = {\rm arg}\{g_{d_\sigma}(\epsilon_k-i\eta)\}$ its phase. The eigenmodes maintain the canonical fermionic anti-commutation relations and are characterized by definite charge and spin/helicity $\sigma$. 

The non-interacting Hamiltonian is expressed in its eigenmodes $\gamma_{k,\sigma}$
\begin{equation}
	{H}_e+\mathcal{H}_t =
	\sum_{\sigma} \int\! dk
	\epsilon_k \gamma^{\dagger}_{k,\sigma}\gamma_{k,\sigma}.
	\label{eq:H_0_scatteringstates}
\end{equation}
They also allow us to write the localized level operators as
\begin{equation}
	d_{\sigma} = \int\! \frac{dk}{\sqrt{2\pi}} 
	t_{\sigma}|g_{d_{\sigma}}(\epsilon_k+i\eta)|
	\gamma_{k,\sigma}.
\end{equation}
The interacting Hamiltonian is then given by \widetext
\begin{align}
  \begin{split}
    \label{eq:H_D_scatteringstates}
	\mathcal{H}_D =& \epsilon_d \sum_{\sigma}t_{\sigma}^2 \int\! \frac{dk dk'}{2\pi} |g_{d_\sigma}(\epsilon_k+i\eta)||g_{d_\sigma}(\epsilon_{k'}-i\eta)| \gamma^{\dagger}_{\sigma,k'}\gamma_{k,\sigma} \\
	&+U t_{+}^2 t_{-}^2 \int\! \frac{dk dk' dq dq'}{(2\pi)^2} |g_{d_+}(\epsilon_k+i\eta)|^2|g_{d_-}(\epsilon_{k'}+i\eta)|^2 \gamma^{\dagger}_{k,+}\gamma_{k',+}\gamma^{\dagger}_{q',-}\gamma_{q,-} \\ 
	&+ \sum_{\alpha,\beta,\nu,\nu'} J_{\alpha,\beta}S^{\alpha} t_{\nu}t_{\nu'}\int\!\frac{dk dk'}{2\pi} |g_{d_\nu}(\epsilon_k+i\eta)||g_{d_{\nu^\prime}}(\epsilon_{k'}-i\eta)| \gamma^{\dagger}_{k',\nu '}\sigma^{\beta}_{\nu',\nu}\gamma_{k,\nu},
  \end{split}
\end{align}
\endwidetext \noindent where we again used $\nu, \nu'$ instead of $\sigma,\sigma'$ in the last term in order to avoid confusion with the notation for the Pauli matrices. Due to the time-reversal symmetry, $t_+=t_-\equiv t$ and $g_{d_{\nu}}(\omega\pm i\eta) = (\omega\pm i\Gamma)^{-1}$ where $\Gamma = \pi\rho_0 t^2$. Note that for this derivation we assumed a wide band limit, $D\gg\Gamma,\omega,eV$, so that the real part of the self-energy of $g_{d_{\nu}}$ can be neglected.

In the limit where $U=0=\epsilon_d$, this Hamiltonian is an anisotropic spin-$1/2$ Kondo Hamiltonian. To see this, we observe that the $J$ term is an exchange coupling between the local spin-density of the $\gamma_{\pm}$ quasiparticles and the local impurity spin $\mathcal{H}_J = \sum_{\alpha,\beta}J_{\alpha,\beta}S^{\alpha}\sigma^{\beta}(0)$ where
\begin{equation}
	\vec{\sigma}(0) = \sum_{\nu,\nu'}\int\!dkdk'
	 \sqrt{\rho_\gamma(k)\rho_\gamma(k')}
	\gamma^{\dagger}_{\nu,k} \vec{\sigma}_{\nu,\nu'} \gamma_{\nu',k'}.
\end{equation}
Here $\pi\rho_{\gamma}(k) = \Gamma|g_d(\epsilon_k)|^2$ is an effective density of states of the $\gamma_{\pm}$ modes that couple to the spin, and $\Gamma$ serves as the bandwidth. In this limit, the setup is characterized by a single Kondo scale $T_K$ for an antiferromagnetic coupling tensor $J_{\alpha,\beta}$. At temperatures below that scale $T\ll T_K$, the local impurity spin will be screened by the $\gamma$-quasiparticles, and the local magnetic moment asymptotically vanish for $T\to 0$ as a Kondo singlet is formed.

As we are mainly interested in the role of the exchange anisotropy on the backscattered current, we will focus first and foremost on the limit where both $U=0$ and $\epsilon_d=0$. We qualitatively discuss how turning them on affects the physics of the setup in subsection~\ref{sec:nozero_epsd_U}.

\subsection{One-loop RG equations and flow}

The advantage of mapping $\mathcal{H}$ onto the Kondo Hamiltonian is
the exploitation of the rich nomenclature and the extensive knowledge of this model. Specifically, the perturbative renormalization group analysis of the Hamiltonian provides already a significant insight into the properties of the setup.

The exchange couplings $J_{\alpha,\beta}$ constitute a tensor, where the first index signifies a component of a vector in the spin space of the quasiparticles $\gamma_{k,\sigma}$ while the second index is a part of a vector in the spin space of the impurity spin. For this section, it will be convenient to write this tensor as comprised of three vectors in the spin-impurity space $\vec{J}_{\beta=x,y,z}$. Each of this vectors is $\vec{J}_{\beta} = \sum_{\alpha}J_{\alpha,\beta}\hat{x}_{\alpha}$ with $\hat{x}_{\alpha}$ being a unit vector in the $\alpha$ direction of the impurity spin. In this notation, $\vec{J}_{\beta}$ couples to the $\beta$ component of the quasiparticles spin density $\sigma^\beta(0)$.

We carry out a poor man's scaling calculation on this setup, in the weak-coupling limit where $|\vec{J}_\alpha|\ll \Gamma$. We relegate the details of the calculations to Appendix~\ref{app:poor_mans} and present and discuss here its results. The RG flow equations close to the local moment fixed point are given by the general expression
\begin{equation}
	\frac{d{\bf J}_{\alpha}}{d\lambda} =
	\pi^2 \rho_0 \epsilon_{\alpha\beta\gamma}
	\vec{J}_\beta \times \vec{J}_\gamma,
	\label{eq:RG_one_loop}
\end{equation}
where $\lambda = \ln(D/D')$ is the logarithm of the running cut-off $D'$.

A detailed analysis of these equations can be found in the appendix of Ref.~\cite{Vinkler-Aviv_2017}. We only present and discuss its main finding here. There are six conserved quantities under this set of equations $a_{\alpha,\beta} = \vec{J}_{\alpha} \cdot \vec{J}_{\beta} = |\vec{J}_{\alpha}| |\vec{J}_{\beta}| \cos(\theta_{\alpha,\beta})$ and $b_{\alpha,\beta} = |\vec{J}_{\alpha}|^2-|\vec{J}_{\beta}|^2$ for $\alpha\neq\beta$. 

For the convenience of the discussion, let us focus now on $a_{x,y}$ and $b_{x,y}$. If $a_{x,y}=0$ and $b_{x,y}=0$ then the coupling is isotropic with $\vec{J}_x \perp \vec{J}_y$ and $|\vec{J}_x|=|\vec{J}_y|$, and the $U(1)$ symmetry is maintained. On the other hand, if $a_{xy}$ and $b_{xy}$ are nonzero, then $U(1)$ symmetry is broken. However, at the strong coupling fixed point $|\vec{J}_{x}|,|\vec{J}_{y}|\to \infty$, from which we can derive
\begin{eqnarray}
	\frac{\vec{J}_x \cdot \vec{J}_y}
	{|\vec{J}_x||\vec{J}_{y}|} &=& 
	\frac{a_{x,y}}
	{|\vec{J}_x||\vec{J}_{y}|}
	 \to 0,
	\nonumber \\ 
	\frac{|\vec{J}_{x}|^2-|\vec{J}_{y}|^2}
	{|\vec{J}_x|^2+|\vec{J}_y|^2} &=& \frac{b_{x,y}}
	{|\vec{J}_x|^2+|\vec{J}_y|^2}
	 \to 0.
\end{eqnarray}
The implication of these limits is that as the magnitude of $|\vec{J}_x|$ and $|\vec{J}_y|$ increase during the RG flow, they flow toward being perpendicular and similar in magnitude. This process describes a dynamical restoration of the $U(1)$ symmetry, and the strong coupling fixed point is isotropic.

We note that not all initial couplings will flow to the strong coupling fixed point, as it is well known that the ferromagnetic Kondo model, with $\vec{J}_x = J_{\perp}\hat{x}$, $\vec{J}_y = J_{\perp}\hat{y}$ and $\vec{J}_z = J_z \hat{z}$ where $J_z < -|J_{\perp}| < 0$, flows to a fixed point where $\vec{J}_{x,y} \to 0$. In this case as well, $a_{x,y}$ and $b_{x,y}$ are zero throughout the entire RG flow, and $U(1)$ symmetry is maintained.

As shown in Ref.~\cite{Vinkler-Aviv_2017}, the backscattering rate is related to the anisotropy and measured by the scale
\begin{equation}
	J_B = \left[\frac{\left(|\vec{J}_x|^2-|\vec{J}_y|^2\right)^2
	+4(\vec{J}_x \cdot \vec{J}_y)^2}
	{\vec{J}_x^2+\vec{J}_y^2}
	 \right]^{1/2}.
\end{equation}
Note that the ${\bf J}_z$ term cannot contribute to the backscattering, since it cannot break the $U(1)$ symmetry. Furthermore, if $\vec{J}_x \perp \vec{J}_y$ and both vectors are of the same length, $J_B=0$.
This defines the line of $U(1)$ symmetric points on which the backscattering current vanishes.
The numerator of $J_B$ is constant under the perturbative RG flow, as it is composed of the conserved $a_{x,y}$ and $b_{x,y}$, while the denominator increases under the flow toward the strong-coupling fixed point. As the low-energy strong-coupling fixed point is isotropic and restores the $U(1)$ symmetry dynamically, we expect the backscattering to vanish when the system reaches that strong coupling fixed point that is beyond the scope of the perturbative RG
analysis.

The formation of the Kondo singlet characterized by the $U(1)$ symmetry is associated with an energy scale $T_K$. In the low-temperature and small bias voltage limit $|eV|, T \ll T_K$, the perfect conductance of the edge will be restored as the backscattering current asymptotically vanishes
for $T\to 0$ and $eV\to 0$. As either the temperature or the bias voltage increases above $T_K$, the RG flow is stopped before the singlet is formed, and the backscattering current may retain a finite value for an initialy $U(1)$ symmetry breaking $\mathcal{H}_D$.

\subsection{Exactly solvable point}

If only one component of the exchange coupling $\vec{J}_{\alpha}$ is nonzero, the interacting problem can be solved exactly. In this case, the projection of $\vec{S}$ parallel to $\vec{J}_\alpha$ is a good quantum number and can be diagonalized together with the Hamiltonian. One implication of only one of $\vec{J}_\alpha$ being nonzero is the absence of any RG flow $d\vec{J}_\alpha/d\lambda = 0$.

As we are interested in exchange coupling that breaks the $U(1)$ symmetry we discuss here the setup where only the component $J_{xx}$ is nonzero. As $S^x$ is a good quantum number, we can diagonalize the Hamiltonian separately for $S^x = \pm 1$. In each sector, the interaction term generates backscattering, where the two sectors are related by time-reversal symmetry.

The fully dressed Green's functions matrices for the $d$ orbitals are given by
\widetext
\begin{align}
  \begin{split}
    \mathcal{G}_d^{r/a}(\omega) &= \frac{1}{(\omega\pm i\Gamma)^2-J_{xx}^2} \left[\begin{array}{cc} \omega\pm i\Gamma & J_{xx}S^x \\ J_{xx}S^x & \omega\pm i\Gamma \end{array}\right], \\
    \mathcal{G}_d^{<}(\omega) &= \frac{2 \Gamma}{|(\omega\pm i\Gamma)^2-J_{xx}^2|^2}
    \left[\begin{array}{cc} (\omega^2+\Gamma^2)f_+ +J^2_{xx}f_- & J_{xx}S^x (\omega+i\Gamma)f_+ + J_{xx}S^x (\omega-i\Gamma)f_- \\ J_{xx}S^x (\omega-i\Gamma)f_+ + J_{xx}S^x (\omega+i\Gamma)f_- & (\omega^2+\Gamma^2)f_-+J^2_{xx}f_+ \end{array}\right],
  \end{split}
\end{align}
with the shorthand $f_{\pm} = f(\omega-\mu_{\pm})$.
\endwidetext

Plugging these expressions into the formula for the current of Eq.~(\ref{eq:I_b_from_G})
and adding the contribution $I_0$ as stated in Eq.\ \eqref{eqn:5}, the full current reads
\begin{equation}
	I(eV) = G_0 V - 4\frac{G_0}{e}J_{xx}^2 \Gamma^2\int\!
	\frac{(f_{+}-f_{-})d\omega}{|(\omega+i\Gamma)^2-J_{xx}^2|^2}.
\end{equation}
At zero temperature the differential conductance approaches
\begin{align}
  \begin{split}
	\frac{1}{G_0}\frac{dI}{dV} = 1- \Bigg[& \frac{2 J_{xx}^2 \Gamma^2}{|(\frac{eV}{2}-i \Gamma)^2-J_{xx}^2|^2} \\
    &+ \frac{2 J_{xx}^2 \Gamma^2}{|(\frac{eV}{2}+i \Gamma)^2-J_{xx}^2|^2} \Bigg] . 
  \end{split}
\end{align}

We note that this is a time-reversal symmetric setup
of the Hamiltonian, where even at zero temperature the zero-bias conductance is not unity and decreases to zero at the point where $|J_{xx}|=\Gamma$. This further illustrates our claim that it is the $U(1)$ symmetry, and not the time-reversal symmetry, that protects the perfect conductance.

\subsection{Nonzero $\epsilon_d$ and $U$}
\label{sec:nozero_epsd_U}
In this section we discuss qualitatively how the previous results are altered when $\epsilon_d$ and $U$ are turned on. The two terms have a significantly different effect. The $\epsilon_d$ term does not affect the results substantially, as it adds a local potential scattering which is marginal in the RG flow, and as long as $\epsilon_d \ll D$, $D$ being the bandwidth, the Kondo singlet will still form as before. At the exactly solvable point discussed above, the addition of $\epsilon_d$ (for $U=0$) is directly incorporated into the Green's functions and the result in that limit is
\begin{equation}
	I(eV) = G_0 V - 4\frac{G_0}{e}J_{xx}^2 \Gamma^2\int\!
	\frac{(f_{+}-f_{-})d\omega}{|(\omega-\epsilon_d+i\Gamma)^2-J_{xx}^2|^2}.
	\label{eq:current_exact_solvable}
\end{equation}

On the other hand, a finite $U$ requires a more delicate discussion. We will separate it into two distinct cases: one without exchange interactions $J_{\alpha,\beta}=0$ and one where the coupling to the impurity spin is turned on.

\subsubsection{The $J_{\alpha,\beta}=0,\; U\neq 0$ case}

Let us first consider the case where the exchange coupling is turned off $J_{\alpha,\beta}=0$ but with finite positive $U>0$. The right and left movers have no way to exchange particles, and $\hat{N}_{\pm}+d^{\dagger}_{\pm}d_{\pm}$ are conserved quantities. The backscattered current is therefore zero regardless of the non-equilibrium conditions. Nevertheless, the physics of this setup are worth discussing.

This is the single impurity Anderson model (SIAM), and for $U>0$ we know that a Kondo peak is created in the density of states of the $d$-orbitals below the Kondo temperature $T^d_K$ and at zero bias if the local orbital occupancy is maintained near integer valence of one: the localized level form a singlet with the conductance electrons. Note that this scale $T^d_K$ differs from the Kondo scale generated by a finite $J_{\alpha,\beta}$.

At finite bias the system is equivalent to a system in equilibrium with a local magnetic field applied to the localized levels as pointed out in Sec.~\ref{sec:symmetry}.\ Note that this is a very subtle point: $\mathcal{H}$ has complex strongly-correlated many-body eigenstates and the many-body scattering states $\Gamma_{k\sigma}$ contain mixtures of right-moving and left-moving edge states. However, the conservation of left and right movers prevents mixing of spin excitations and the non-equilibrium $Y$ operator maintains the form of a Zeeman term, as $[\mathcal{S}^z, \mathcal{H}]=0$ still holds.
Time reversal symmetry is maintained by the Hamiltonian and only broken
by the externally applied bias that enters in the $Y$ operator and drives the edge 
current $I$.

\subsubsection{Finite $U$ and weak $J_{\alpha,\beta}$}

We also briefly consider the case where both $U>0$ and $J_{\alpha,\beta}$ are finite. Starting from $J_{\alpha,\beta}=0$ and $\epsilon_d+U>0$, the Hamiltonian approaches the strong coupling fixed point~\cite{Bulla2008} below  $T^d_K$. This fixed point describes a local Fermi liquid that can be treated as a free electron gas for $|\omega|<T_K^d$ in leading order. Switching on an anti-ferromagnetic  $J_{\alpha,\beta}$ leads to another Kondo effect~\cite{Lebanon_2003}
involving the screening of the local spin below the temperature $T^s_K$ that is exponentially dependent on the $J_{\alpha,\beta}$~\cite{Wilson1975}.
This picture remains valid for $T_K^s\ll T_K^d$ and generates a pseudo-gap in the full renormalized orbital spectral function $\rho_d(\omega)$.

We derive an expression for the backscattered current by treating $J$ perturbatively and then follow a similar approach to the one taken above for the exactly solvable point. Since a $U(1)$ symmetric $J_{\alpha,\beta}$ leads to vanishing $I_B$ we restrict ourselves to a finite $J_{xx}$ term and set all other $J_{\alpha,\beta}=0$. In leading order in $J_{xx}$, backscattering happens between the two local Fermi liquids. The backscattered current will therefore be
\begin{equation}
	I_{B} \sim \frac{G_0}{e}J_{xx}^2\int\! 
	\rho_{+}(\omega) \rho_{-}(\omega) (f_+-f_-) d\omega,
\end{equation}
where $\rho_{\pm}(\omega)$ is the renormalized density of states of $d_{\pm}$, including the effects of $t,\epsilon_d$ as well as $U$. We wrote $J_{xx}$ for connection with the formula in Eq.~(\ref{eq:current_exact_solvable}), but one has to sum over all terms $J_{\alpha,\beta}$ that allow backscattering.

Let's assume we have finite $U$ and $J$ and two Kondo scales $T_K^d$ and $T_K^s$. If $J$ is large then the local spin and the d orbital will form a 
singlet and decouple from the edge. For small $J$, the argumentation
above holds and the d orbital will get screened at first and then
screen the local spin in turn. This leaves us with two distinctly
different GS for small and large $J$.
The parameter space of weak and strong $J_{\alpha,\beta}$ and a finite $U$
are adiabatically connected: Since  there is no quantum phase transition in the parameter space we leave the analysis of the full parameter space where both $J_{\alpha,\beta}$ and $U$ are finite and comparable to a later study.
Here, we are interested on the fundamental mechanism generating 
a backscattered current $I_B$ in a time-reversal symmetric Hamiltonian.
From now on we mostly discuss the case where $U=0$, which will allow us to focus on the role of the exchange coupling anisotropy. In this case, the Kondo temperature $T_K^d$ is maximal and replaced by $\Gamma$. Therefore we always choose the parameters for the numerical simulation such that $T_K^s <\Gamma$.

\section{Numerical analysis}
\label{sec:td_NRG}
\subsection{Time-Dependent Numerical Renormalization Group and Green's
Functions}
The backscattering current Eq.~\eqref{eq:I_b_from_G} requires calculation
of the non-equilibrium retarded and lesser Green's functions. Since we
are interested in the low-temperature behavior for arbitrary
interaction strength $J_{\alpha\beta}$ as well as a wide range of bias
voltages, we opt for the TD-NRG~\cite{Anders2005, Anders2006}, which has been
used successfully to calculate steady-state Green's Functions in
the context of transport through single-orbital quantum dots before
\cite{Anders2008, Anders2010}. 
It also allows to access the low-energy fixed point in equilibrium  of
the model introduced in Sec.\ \ref{sec:perturbative_RG} and, therefore, test the predictions of the analytical perturbative RG approach outlined above.

The NRG
was originally developed by Wilson~\cite{Wilson1975} to solve the
single-channel Kondo problem but has been extended to 
various problems describing magnetic impurities coupled to a host's
conduction bands in the meantime. The general Hamiltonian, as discussed
in Sec.~\ref{sec:model_and_current}, can be
partitioned into three parts
\begin{align}
\mathcal{H} = \mathcal{H}_{D} + \mathcal{H}_{e} + \mathcal{H}_{t},
\end{align}
where $\mathcal{H}_{D}$ and $\mathcal{H}_{e}$ contains
impurity or edge degrees of freedom only. The impurity part
may comprise local many-body interactions of arbitrary strength. The edge states,
however, are taken to be non-interacting and play the role of the
quasi-continuous band in the conventional NRG.\@ The third term
$\mathcal{H}_{t}$ describes a hybridization between the localized
impurity and the edge states. In the NRG scheme, one proceeds by
partitioning the hybridization function $\Gamma(\omega)$ into
logarithmically shrinking intervals around the chemical potential with
help of the dimensionless discretization parameter $\Lambda>1$. The
edge degrees of freedom are rewritten as linear combinations of
operators for each such interval. Only modes that couple directly to
the impurity are retained at this point. The system is further
transformed by a tridiagonalization algorithm and mapped onto a
semi-infinite tight-binding chain, the so-called Wilson chain, where the
first chain link is equivalent to $\mathcal{H}_{D}$. The system is
now solved in an iterative fashion where one diagonalizes the
Hamiltonian for a given chain length, calculates expectation values of
interest, and proceeds by adding the next chain link. The
tight-binding hopping parameters of such a chain fall off exponentially as one
traverses the chain which is a direct result of the logarithmic discretization
of the hybridization function. Due to the
exponentially decreasing hopping elements, the Hamiltonian of a given
iteration can be linked to a likewise decreasing temperature
scale~\cite{Wilson1975, Bulla2008}. The iterative scheme is terminated at some finite chain length
$N$ that determines the target temperature $T_N \sim \Lambda^{-N/2}$.

Only the $N_s$ states
with the smallest eigenvalues are kept each iteration and coupled to the next chain
link in order to tackle the otherwise exponentially growing Fockspace.
Furthermore, we employ Oliveira's $z$ averaging~\cite{Oliveira1994} to suppress
discretization artifacts and improve numerical precision. 

In the TD-NRG~\cite{Anders2005, Anders2006}, we regard the system to
be in thermal equilibrium for 
$t < 0$, at which point an additional interaction term $\Delta \mathcal{H}$ is
turned on. Thus, the Hamiltonian undergoes an abrupt change (or quench)
\begin{align}
  \mathcal{H}^{i} \rightarrow \mathcal{H}^{f}(t>0) = \mathcal{H}^{i} + \Delta \mathcal{H}.
\end{align}
As a result, the density operator for $t>0$ evolves in time with
respect to the final Hamiltonian $\mathcal{H}^{f}$
\begin{align}
  \rho(t>0) = \mathrm{e}^{-\mathrm{i} \mathcal{H}^{f} t} \rho_0 \mathrm{e}^{\mathrm{i} \mathcal{H}^{f} t}.
\end{align}
The equilibrium NRG scheme described above needs a further
refinement since
non-equilibrium calculations involve contributions from all energy
scales intermingled together. One can show~\cite{Anders2005,
  Anders2006} that a set of all discarded states form a complete basis
for a Wilson chain of length $N$. Conceptually, one first carries out two separate equilibrium NRG
calculations for $\mathcal{H}^{i}$ and $\mathcal{H}^{f}$
respectively. The eigenbasis of the final Hamiltonian is needed for
the time-evolution of any operator $O(t)$ while the reduced density matrix is
constructed in the eigenbasis of the initial Hamiltonian. The overlap
matrix $S_m$ allows for rotation between both bases at given iteration
$m$ and connects both NRG runs.

The approach outlined above can be straightforwardly extended for
equilibrium spectral functions in their Lehmann
representation~\cite{Peters2006, Weichselbaum2007}. The TD-NRG and the 
sum-rule conserving scheme for the spectral functions were combined in
Ref.~\cite{Anders2008_NEQGF} to evaluate non-equilibrium
Green's functions for times $t,t^\prime$. Note that both, the equilibrium and
the non-equilibrium calculations, can be extended readily to lesser
and greater Green's functions~\cite{Weichselbaum2007,
  Anders2008_NEQGF}. The spectral $\delta$-functions of the Lehmann representation are
broadened by a logarithmic Gaussian as defined in Eq.\ (74) in Ref.~\cite{Bulla2008},
where we used the broadening parameter ${b=0.8}$ throughout the paper.

Evaluation of the backscattering current Eq.~\eqref{eq:I_b_from_G}
poses a number of challenges from a technical point of view. First,
the calculations of the non-equilibrium Green's functions themselves
according to the TD-NRG procedure. Second, we are not able to employ
the improvement of the NRG Green's function via an equation of
motion~\cite{Bulla1998} since it is not readily applicable for
non-equilibrium lesser Green's functions. Third, we need to calculate
a difference between retarded spectral function and lesser Green's
function, that may well be very small, before integrating numerically
over the whole real axis. Finally, we are interested in the linear
conductance $G=I_B / V$ which limits our precision further and keeps
us effectively from using arbitrary small bias voltages since the
already small current $I_B$ cannot be distinguished from numerical
noise in the limit $V\to 0$.

In the following we choose a discretization parameter $\Lambda=2$, a
half-bandwidth $D/\Gamma=100$, and $z$ averaging of $z=4$ for all our
calculations. If not stated otherwise, we use a Wilson chain of length
$N=45$ which results in a target temperature $T/\Gamma\approx
1.79\cdot10^{-5}$. This choice of parameters guarantees
that the temperature $T$ for our calculations is well below
the equilibrium Kondo temperature $T_K^{\rm eq}$ as we will discuss in the next section.

\subsection{Equilibrium and effective equilibrium}
We start by addressing the setup in equilibrium. While
we are mostly interested in the case where $U=0$ and
$J_{\alpha,\beta}\neq 0$, it is instructive to first consider the
opposite case where $U$ is finite and $J_{\alpha,\beta}$ are turned
off. Under this conditions, the additional spin completely decouples
from the subsystem comprising the local $d$ orbital and the edges, and
the system is equivalent to an equilibrium Single Impurity Anderson Model (SIAM). 

\begin{figure}[tbp]
  \includegraphics[width=0.5\textwidth,clip]{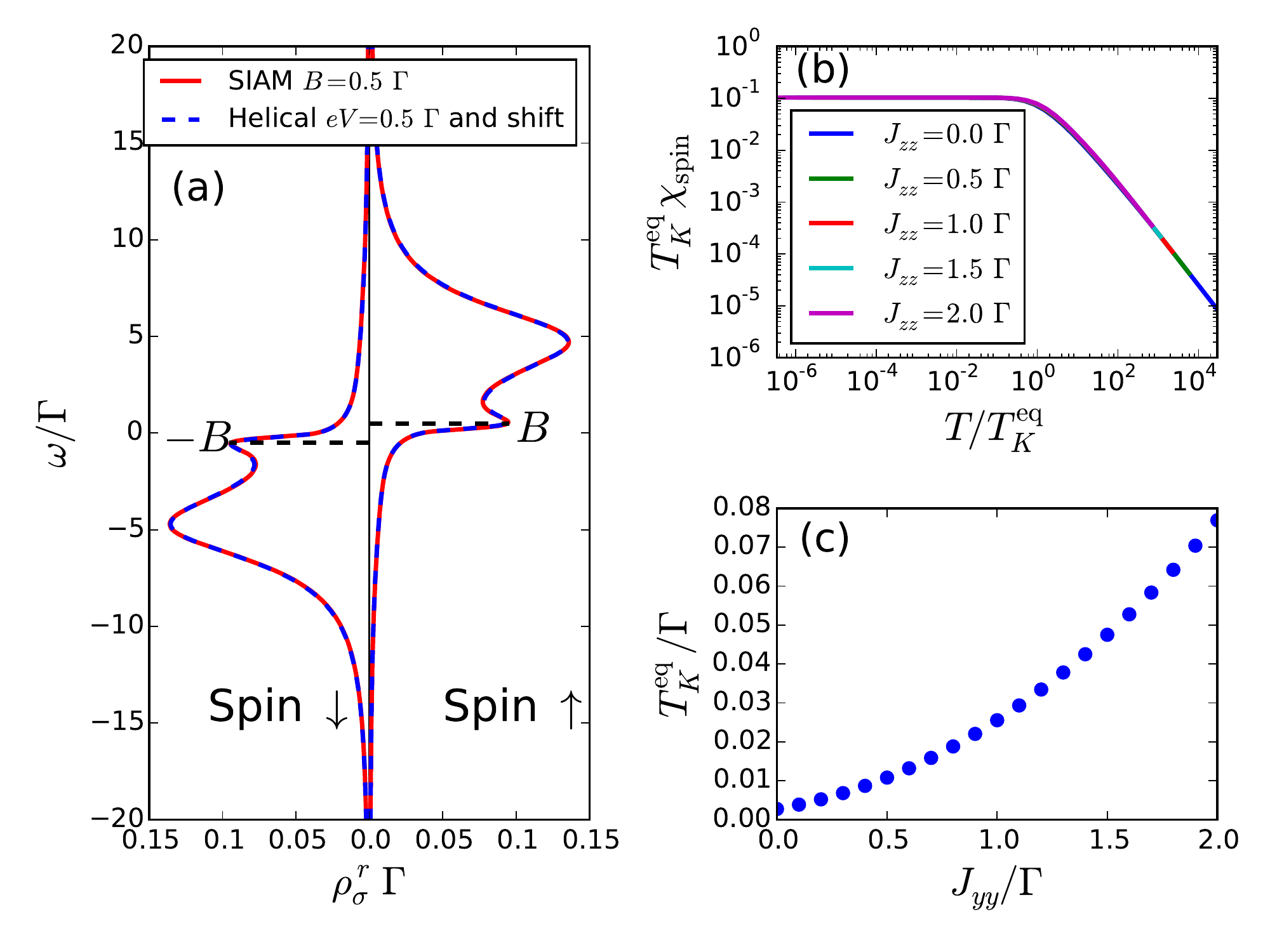}
  \caption{\label{fig:equilibrium} (a) $\rho_\sigma^{r,\mathrm{SIAM}}$
    (red solid line) for finite magnetic field $B/\Gamma =0.5$
    compared to $\rho_\sigma^{r,\rm helical}$ (blue dashed line)
    calculated for bias voltage $eV/\Gamma=0.5$ and
    $J_{\alpha\beta} = 0$. The results for the helical model are
    shifted by an additional $\pm eV/2$. For both models
    $\epsilon_d/\Gamma =-5$ and $U/\Gamma =10$. (b) Local spin
    susceptibility $\chi_{\rm spin}$ for helical model in equilibrium
    $eV=0$, $\epsilon_d/\Gamma=-0.5, U=0$, and
    ${J_{xx}=J_{yy}=\Gamma}$. (c) Equilibrium Kondo temperature
    calculated from the local spin susceptibility as a
    function of $J_{yy}$ for ${J_{xx}=J_{zz}=\Gamma}$ and
    ${U=\epsilon_d=0}$.}
\end{figure}

We performed two independent NRG calculations: a conventional equilibrium
NRG calculation of a SIAM in a finite magnetic field, and a full scattering states TD-NRG
calculations where the bias enters through the $Y$ operator in the density
operator but the dynamics is governed by  the Hamiltonian only
\cite{Anders2008_NEQGF}.
Remarkably,
as discussed above, the system remains in effective equilibrium even
when a finite bias voltage is applied, as the two spin-flavors are
only capacitively connected. When a bias voltage $eV$ is applied, the
system behaves as under the influence of a magnetic field $B$ where
the chemical potential difference takes on the role of the Zeeman
energy. Here, the Kondo peak resides at $\pm B$ for 
spin up and spin down, thus accounting for a splitting of $\Delta
E = g_{\rm eff} B$ while the Kondo peak forms around the respective
chemical potential in the helical model. As a result, the spectral density
of the equilibrium SIAM calculation $\rho_\sigma^{r,\mathrm{SIAM}}$ shows a peak at
double the chemical potential of the opposite spin on an absolute
scale [Fig.~\ref{fig:equilibrium}\,(a)]. Perfect agreement can be
realized by a symmetric shift of $\pm eV/2$.

We are ultimately interested in the backscattered current driven by
finite exchange coupling to the local spin. In order to examine the
role of the anisotropy, we turn on a finite $J_{\alpha,\beta}$ and set
$U=0$. The finite $U$ regime is adiabatically connected but results in
a much lower characteristic energy scale.  In equilibrium $eV=0$, this
setup is also characterized by a Kondo screening, which is different
than the Kondo screening for the SIAM setup (finite $U$ and zero
exchange coupling) discussed before. The Kondo temperature associated
with this exchange coupling can be found numerically by employing
Wilson's definition using the temperature dependent magnetic
susceptibility via
${4T_K^{\rm eq}\chi_{\rm spin}(T_K^{\rm eq}) =
  0.413}$~\cite{Wilson1975, Bulla2008}. Here, $\chi_{\rm spin}(T)$ is
calculated by applying an infinitesimal small local magnetic
field and measuring the polarization of the localized spin (not the
spin of the $d$ electron) in absence of a bias voltage ${eV=0}$
[Fig.~\ref{fig:equilibrium}\,(b)]. In the following, we will refer to
the equilibrium Kondo temperature calculated in this way as
$T_K^{\rm eq}$ to emphasize that it stems from an equilibrium
calculation. To simplify the discussion, we restrict ourselves to
exchange couplings that contain only diagonal terms
$J_{\alpha,\alpha}$. We note that it is sufficient to tune the ratio
${J_{xx}/J_{yy}}$ to break $U(1)$ symmetry and generate a
backscattered current, as discussed above
(Sec.~\ref{sec:symmetry}). This has the added benefit of eliminating
complex terms from the local Hamiltonian, simplifying the numerical
calculations. We also take advantage of the fact that $J_{zz}$ does
not affect the $U(1)$ symmetry, and we can set it at will. For the
$U(1)$ symmetric point where $J_{xx}=J_{yy}=J_{zz}=\Gamma$, we get an
equilibrium Kondo temperature ${T_K^{\rm eq}/\Gamma\approx 0.025}$
[Fig.~\ref{fig:equilibrium}\,(c)].

\subsection{Finite backscattered current for $eV>T_{K}^{\rm eq}\gg T$}
\label{subsec:cutoff}
\begin{figure}[tbp]
  \includegraphics[width=0.5\textwidth,clip]{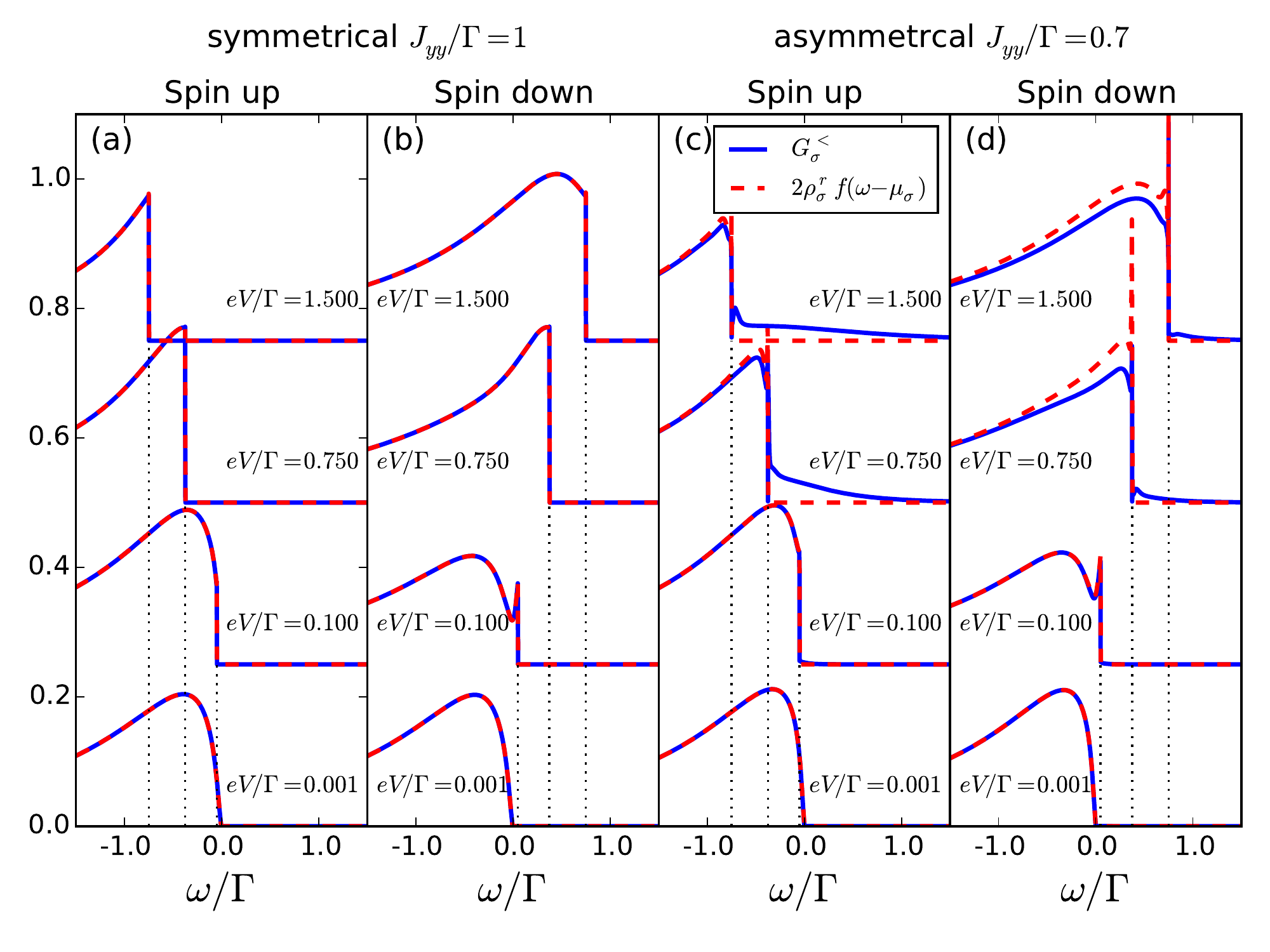}
  \caption{\label{fig:greens-function}${G^<_\sigma(\omega)}$ (solid blue curves)
    and ${2\rho_\sigma^r(\omega) f(\omega - \mu_\sigma)}$
    (dashed green curves) for (i) the symmetrical point ${J_{yy}/\Gamma =
      1}$ for spin (a) up and (b) down and (ii) for the broken $U(1)$ symmetry,
    ${J_{yy}/\Gamma = 0.7}$, for spin (c) up and (d) down.
    $G^<_\sigma$ and $\rho_\sigma^r$ for consecutive bias
    voltages $eV$ are shifted by a constant offset $a = 0.25$ for
    better visibility. The legend applies to all subplots.
  }
\end{figure}

We start at the symmetrical point ${J_{xx} = J_{yy} = \Gamma}$ and
turn on a gate voltage $eV$ on the edge. Below, we quantify the
deviation from the $U(1)$ symmetric point by
$\Delta J_{yy} = J_{yy} - \Gamma$ and retain the other two exchange
parameter at fixed values $J_{xx} = J_{zz}= \Gamma$. The problem thus
becomes a full non-equilibrium one.  Both the lesser Green's
  function (GF) $G_\sigma^<$ and the spectral function $\rho_\sigma^r$
  times Fermi function fall off at the chemical potential for the
  respective spin $\sigma$. In the symmetrical case, the system can
be mapped to an effective equilibrium problem and the lesser GF is
equal to the retarded spectral function times the Fermi function and
appropriate constant factor as a direct consequence of the
fluctuactions-dissipations-relation
[Fig.~\ref{fig:greens-function}\,(a) and (b)]. We break the $U(1)$
symmetry by performing a quench in the value of $\Delta J_{yy}$. In the
asymmetrical case and for ${eV>T_K^{\rm eq}}$, the non-equilibrium
lesser GF and retarded spectral density start to differ
[Fig.~\ref{fig:greens-function}\,(c) and (d)] which consequently
drives a backscattering current. The NRG GF broadening induces small
finite size oscillations~\cite{Bulla2008} in the spectral functions at
the chemical potentials and the numerical integration. This
effectively limits our precision for the backscattered
conductance calculated by the integral over the difference
between both GFs.

The conductance can be partitioned into two regimes: (i) ${eV<T_K^{\rm
    eq}}$ and (ii) ${eV>T_K^{\rm eq}}$ which are connected by a
crossover regime. In both cases we consider the temperature being
the smallest energy scale i.\ e.\ ${T\ll T_K^{\rm eq}, eV}$.
For bias voltages that are lower than $T_K^{\rm eq}$, the
system cross-over to a regime in which the impurity spin is screened
and $U(1)$ symmetry is dynamically restored. As a consequence, the
backscattered current vanishes even when the initial parameters
break the $U(1)$ symmetry, implying that the total edge has a perfect
zero-bias differential conductance.

\begin{figure}[tbp]
  \includegraphics[width=0.5\textwidth,clip]{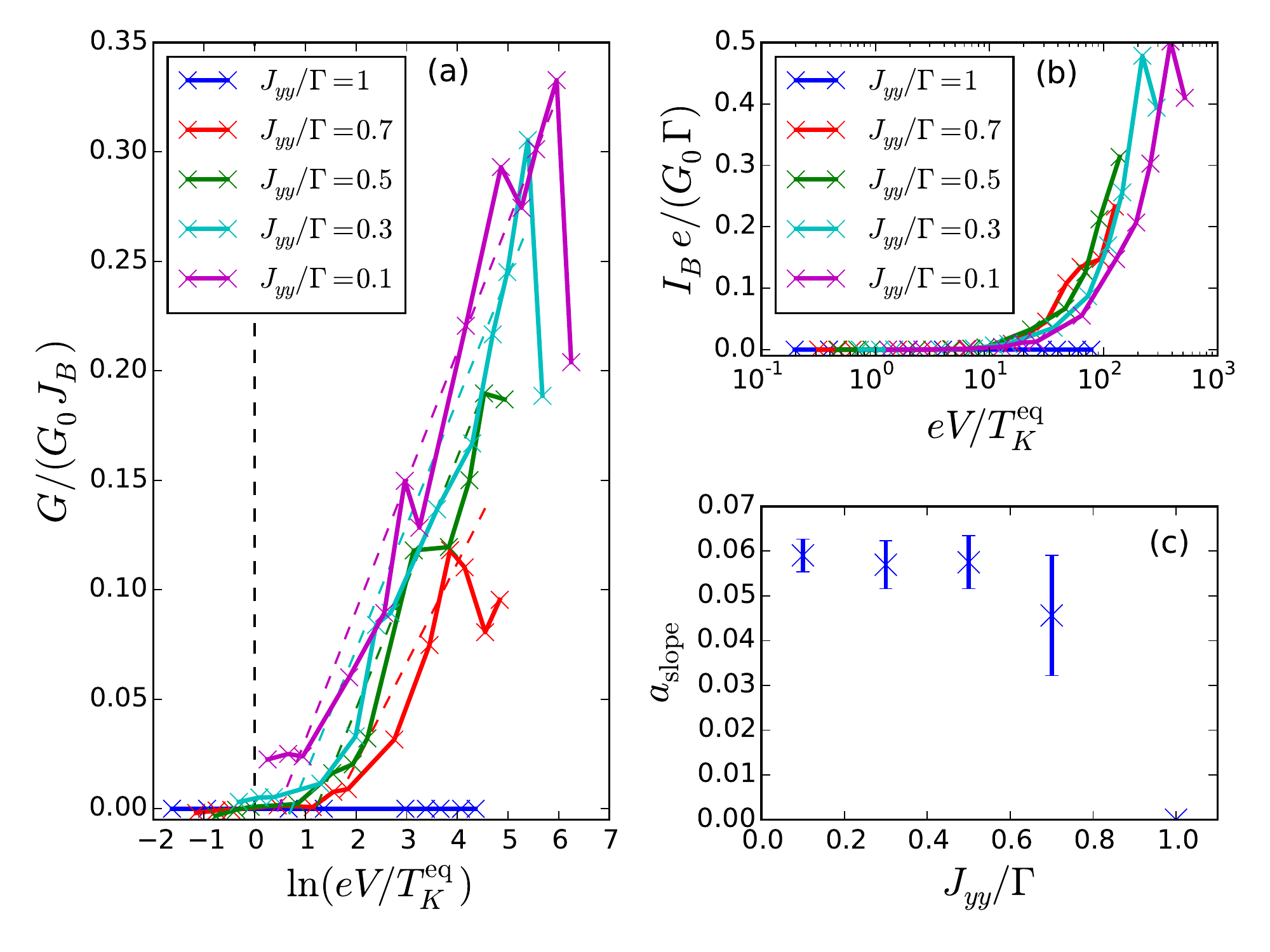}
  \caption{\label{fig:cutoff} (a) Linear conductivity ${G=I_B/V}$ as
    function of voltage ${eV/T_K^{\rm eq}}$ for different couplings
    ${J_{yy}}$. $I_B=0$ for the $U(1)$ symmetric case
    $J_{yy}=\Gamma$. The vertical black dashed line indicates
    ${eV=T_K^{\rm eq}}$. The other dashed lines represent the fits to
    Eq.\ \eqref{eq:35-log-fit}. (b) Backscattering current $I_B$ as
    function of ${eV/T_K^{\rm eq}}$. (c) Slopes $a_{\rm slope}$ for the
    different fits in subplot (a). The error bars stem from 
    the numerical fitting process. The values for ${T_K^{\rm eq}}$ are
    shown in Fig.~\ref{fig:equilibrium}\,(c). In all cases
    $J_{xx}=J_{zz}= \Gamma$.
  }
\end{figure}

For a setup with broken $U(1)$
symmetry, the equilibrium RG flow equations \eqref{eq:RG_one_loop} are
cut-off by ${eV>T_K^{\rm eq}}$~\cite{BordaZawa2010,
  RoschKrohaWoelfe2001}, therefore preventing the system from approaching
the strong coupling fixed point and restoring the perfect edge. In the
symmetric case, $\Delta J_{yy}=0$, the fluctuations-dissipations theorem holds perfectly
for each spin sector individually, and the conductance
vanishes regardless of $eV$. 

Numerically we find small
negative values for $I_B$ in the ${eV<T_K^{\rm eq}}$ regime for broken symmetry
that we trace back to three sources of errors. Firstly, the error
increases with increasing the quench $\Delta J_{yy}$ as a consequence
of the discrete representation of the continuum problem by the Wilson
chain~\cite{Anders2006, EidelsteinGuettgeSchillerAnders2012, GuettgeAndersSchiller2013}. 
Secondly, the smaller $eV$ the smaller the difference between both GFs
will be indicated in Fig.~\ref{fig:greens-function}. Therefore, the
relative error due to subtraction and integration is increasing.
Thirdly, the linear conductance is proportional to
$V^{-1}$ requiring a high numerical precision of the integral
determining $I_B$ for small $eV$ in Eq.~\eqref{eq:I_b_from_G}. A
voltage of order ${eV/\Gamma\sim 10^{-3}}$ demands a precision of the backscattering
current of at least four relevant digits. Here, not only the scattering states NRG 
but also the  discretization of the spectral function on a finite frequency grid 
generates a small error in the numerical integration. We find that the smallest
voltage, for which we could still get results that are not
overshadowed by numerical noise, is ${eV/\Gamma \approx 0.005}$.

When we start in the large $eV$ regime and decrease the voltage,
the finite $G$ for broken symmetry is also reduced until the system reaches
the small $eV$ regime. In the crossover regime, we
extracted the parameter of a fitting function
\begin{align}
  \label{eq:35-log-fit}
  G/J_B = a_{\rm slope}\ln(eV/T_K^{\rm eq}) + b
\end{align}
to the data shown in Fig.~\ref{fig:cutoff}(a). The function is added as 
dashed lines in the same color to the plot. We find
that the slope $a_{\rm slope}$  depicted in Fig.~\ref{fig:cutoff}(c)
is nearly independent of the coupling constant $J_{yy}$. 
  
\subsection{Finite backscattered current for $T>T_{K}^{\rm eq}$}
\label{subsec:T_dependence}
\begin{figure}[tbp]
  \includegraphics[width=0.5\textwidth,clip]{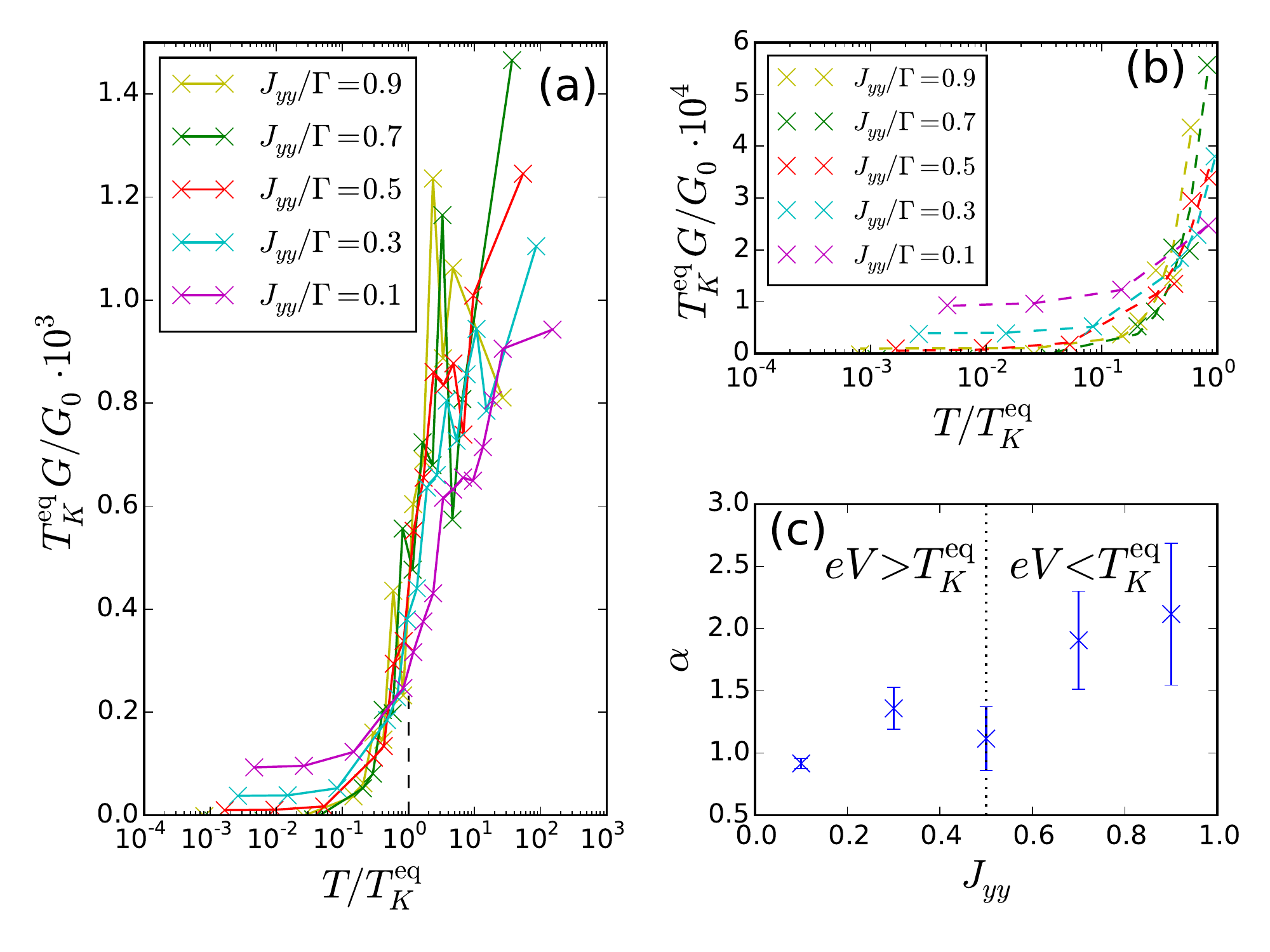}
  \caption{\label{fig:T-evolution}(a) Linear conductivity $G$ as function
    of temperature ${T/T_K^{\rm eq}}$ for a fixed ${eV/\Gamma=0.01}$ and
    different couplings $J_{yy}$. The black dashed line indicates
    $T_K^{\rm eq}$. Below ${T/T_K^{\rm eq} = 1}$ the conductance $G\to
    0$ when ${eV< T_K^{\rm eq}}$ (yellow, green curve). The lines are a
    guide to the eye. (b) Power-law fit (Eq.\ \eqref{eq:36-power-law})
     to the data points (crosses) of (a) for $T/T_K^{\rm eq} <1$.
    (c) Exponent $\alpha$ of the power-law fit: $\alpha \to 2$
    (Fermi-Liquid) for ${eV< T_K^{\rm eq}}$.  The error bars stem from
    the numerical fitting process.
  }
\end{figure}

As in the previous section, we retain the parameters of a diagonal
matrix $J_{\alpha\beta}$ with $J_{xx}=J_{zz}=\Gamma$ and use $J_{yy}$
as a tuning parameter. The cut-off of the RG flow-equations does not
necessarily have to come from high bias voltage but can be due to
finite temperature as well. For the regime $eV<T_K^{\rm eq}<T$, we
expect that a setup with a broken $U(1)$ symmetry will not have its
edge reconstructed and a finite backscattering will be observed. We
choose a fixed voltage ${eV/\Gamma = 0.01}$ and calculate $G$ as
function of $T$ for various couplings $J_{yy}$ in the symmetry broken
regime. Our particular choice partitions our results into two groups:
for ${J_{yy}/\Gamma\ge 0.7}$, we find ${T_K^{\rm eq} > eV}$ while the
voltage is the largest energy scale for ${J_{yy}/\Gamma\le 0.3}$. For
${J_{yy}/\Gamma= 0.5}$, Kondo temperature and voltage are almost equal
and the system is located in the crossover regime.

In the first case, the low temperature conductance shows a universal
behavior for ${T<T_K^{\rm eq}}$ approaching asymptotically zero
[Fig.~\ref{fig:T-evolution}\,(a)], as discussed in the previous
section. If $eV$ is the largest energy scale, then the conductance
converges towards a finite value for $T\to 0$. This asymptotic value
increases monotonically with the ratio $eV/T_K^{\rm eq}$ (see
Fig.~\ref{fig:T-evolution}\,(a) cyan and magenta curve), i.e. the
earlier the perturbative RG flow equations are cut off by $eV$.

The low temperature behavior of $G$ is converged and, in case of
${eV<T_K^{\rm eq}}$, is expected to follow a power-law. We use a fit of
the form
\begin{align}
  \label{eq:36-power-law}
  T_K^{\rm eq} G (T) = b (T/T_K^{\rm eq})^{\alpha} + c
\end{align}
and determine the exponent $\alpha=2$ from the data presented in
Fig.~\ref{fig:T-evolution}\,(b) as depicted on the right side of
Fig.~\ref{fig:T-evolution}\,(c). An exponent of $\alpha=2$ is associated with single-particle backscattering, but as noted in previous studies~\cite{Vayrynen_2014, Lezmy_2012} the nature of the low-energy fixed-point Hamiltonian is strongly restricted by symmetry considerations, and cannot contain a single-particle backscattering term as such a term will break time-reversal symmetry. While maintaining time-reversal symmetry, the leading possible perturbation is a two-particles backscattering term which should have a power-law corresponding to $\alpha=4$. However, by applying finite voltage we, and any experimental setup, explicitly break time-reversal symmetry, and we understand the strong $\alpha=2$ exponent to be a signature of non-equilibrium physics, with $G \propto (eV T)^2/T_K^4$. This suggests that for finite voltages the anisotropy might be the most dominant cause for the deviation from perfect conductance of the edge that was observed in experiments~\cite{Jia_2017,Fei_2017,Sabater_2013,Mueller_2015,Li_2017,Suzuki_2015, Du_2015,Spanton_2014,Suzuki_2013, Olshanetsky_2015,Knez_2011,Gusev_2013,Gusev_2014a,Gusev_2014b, Grabecki_2013,Nowack_2013,Kononov_2015,Brune_2012,Roth_2009,Konig_2007}.

The offset $c$ in Eq.~\eqref{eq:36-power-law} is numerically zero in the regime where ${eV<T_K^{\rm eq}}$. In case $eV$ is the relevant low-energy scale, it attains a nonzero value $c=c(eV/T_K^{\rm eq})$, which is not constant and increases with $eV$.

\subsection{Dynamical restoration of the $U(1)$ symmetry and the breakdown of backscattering current}
\label{subsec:breakdown_backscattering}
\begin{figure}[tbp]
  \includegraphics[width=0.5\textwidth,clip]{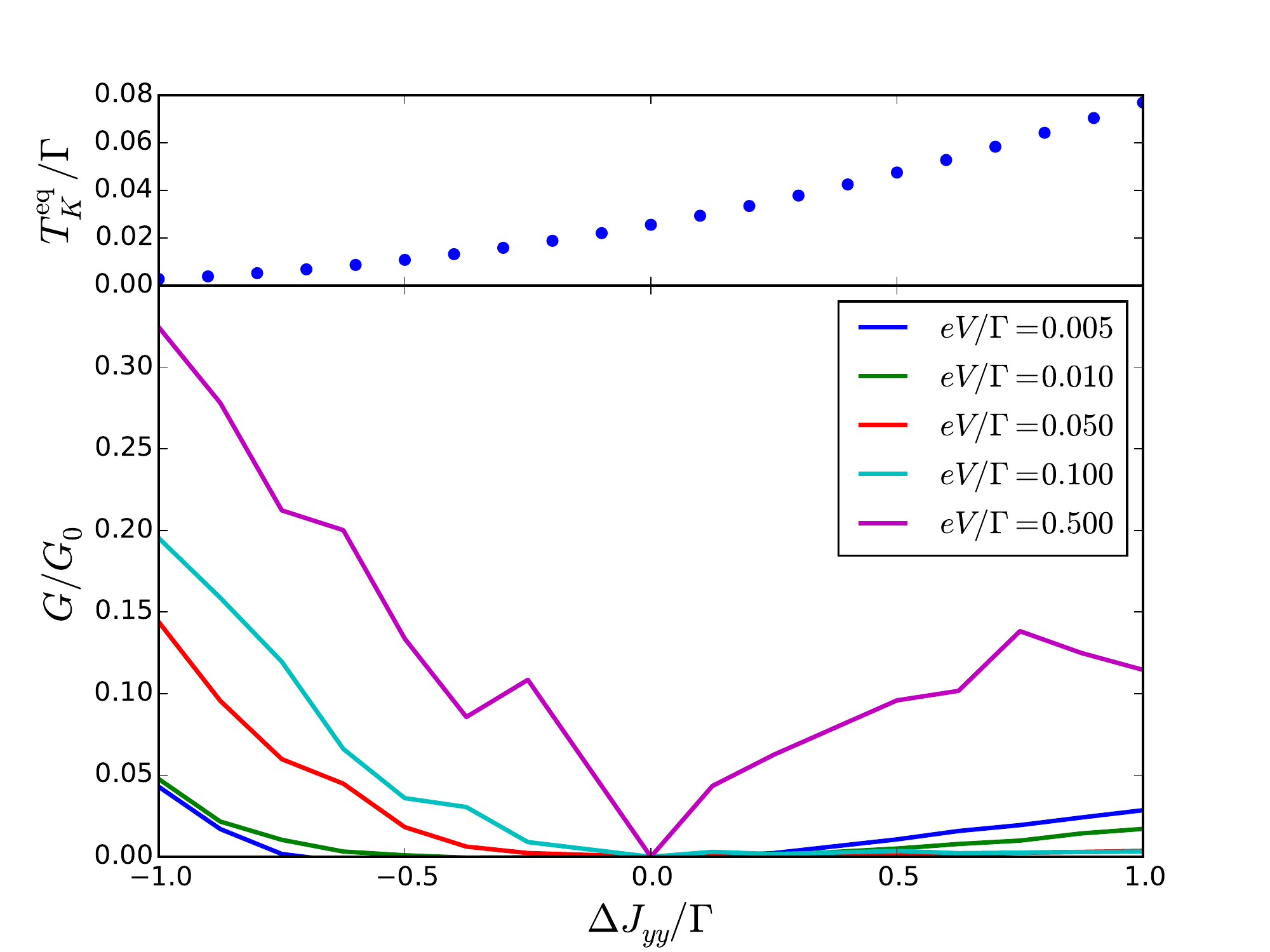}
  \caption{\label{fig:V-shape}Low temperature $G=I_B/V$ as a function of $\Delta J_{yy}$ 
    for different bias voltages $eV$ and $T\to 0$. The upper subplot shows the
    corresponding equilibrium Kondo temperatures $T_K^{\rm eq}(\Delta
    J_{yy})$. 
  }
\end{figure}

Now we focus on the behavior of the conductance in the limit
  $T\to 0$ and finite $eV$ as we break the $U(1)$ symmetry by a
finite detuning $\Delta J_{yy}=\Gamma-J_{yy}$ and holding all other
parameters fixed. The conductance vanishes at the symmetrical point
$\Delta J_{yy}=0$ (${J_{yy}/\Gamma = 1}$) regardless of $eV$. The
symmetric point is asymptotically restored by the Kondo effect in the
limit $T\to 0$. Note, that the corresponding Kondo temperature
$T_K^{\rm eq} = T_K^{\rm eq}(J_{yy})$ depends on the exchange coupling
for otherwise fixed parameters. We calculate the conductance $G$ as
function of $\Delta J_{yy}$ for a fixed bias voltage $eV$ in the limit
$T\to 0$ and plot the curve for different $eV$ in
Fig.~\ref{fig:V-shape}. If $U(1)$ symmetry is broken, the conductance
depends on the ratio ${eV/T_K^{\rm eq}}$ as discussed above.

When $T_K^{\rm eq}$ replaces $eV$ as the largest low energy scale, the
strong coupling fixed point is approached and backscattering is
suppressed explaining the vanishing conductance for ${J_{yy}\Gamma<1}$
in Fig.~\ref{fig:V-shape}. When $T_K^{\rm eq} \approx eV$, a
backscattering current is found as shown for large negative
$\Delta J_{yy}$.

For ${J_{yy}/\Gamma>1}$, $T_K^{\rm eq}$ is generally the largest
energy scale except for ${eV/\Gamma=0.5}$ (magenta curve) which still
shows a significant backscattering conductance. The conductance is
smaller than for ${J_{yy}/\Gamma<1}$ due to the higher
${T_K^{\rm eq}}$ as the renormalization process is cut-off later and
the system moves closer to a strong coupling fixed point. We find a
finite conductance for large $J_{yy}$ and ${eV/\Gamma<0.01}$ albeit
${eV<T_K^{\rm eq}}$. We again attribute this residual $G$ to numerical
inaccuracies in the calculation of the current via
Eq.~\eqref{eq:I_b_from_G}. The calculation of this residual $G$
requires an accuracy of 5 digits at $eV/\Gamma<0.01$ which is beyond
our numerical precision. We conclude that $G\to 0$ for
$eV/\Gamma<0.01$. We believe that once $eV$ exceeds $0.1\Gamma$ and
$eV$ of $O(T_K^{\rm eq})$, the backscattering current emerges from the
numerical noise in the regime $\Delta J_{yy}>0$.

In short, we found a vanishing backscattering current at the symmetry
point. The renormalization process is cut off for  ${eV>T_K^{\rm eq}}$,  
and a finite $I_B$ remains for broken $U(1)$ symmetry. 

\section{Summary and Discussion}
\label{sec:discussion}
In this paper we studied and analyzed the conductance of helical edge
modes when coupled to a magnetic impurity, combining analytical and
numerical methods. We derived a general expression for the
  non-equilibrium dc-current in Eq.~\eqref{eq:I_b_from_G} by coupling
  the helical edge electrons to localized levels. The current is
  independent of the specific details of the interactions of the local
  levels, which are encoded implicitly in the Green's functions for
  the localized levels. An
analysis of the expression for the current using time-dependent gauge
transformations as well as Hershfield's formalism revealed the role of
a global $U(1)$ symmetry in protecting the perfect conductance of the
helical modes. If the $U(1)$ symmetry is retained, then the edge
manifests a perfect conductance even if time-reversal symmetry is
broken. This conclusion was further corroborated by considering a
specific exactly solvable interacting setup that maintains
time-reversal symmetry but breaks $U(1)$ symmetry. We demonstrated in
Eq.~(\ref{eq:current_exact_solvable}) that the conductance is not
perfect even at zero-temperature and zero-bias. Similarly, the case
where $U(1)$ symmetry was preserved but time-reversal symmetry broken
was mapped onto an equilibrium setup with perfect conduction.

We then focused on an interaction Hamiltonian consisting of an
exchange coupling between the levels and a localized impurity spin,
defined by the coupling tensor $J_{\alpha,\beta}$ which allows for
anisotropies that break the $U(1)$ symmetry. The one-loop RG flow
equations of the exchange coupling, given in
Eq.~(\ref{eq:RG_one_loop}), showed that in general there is a
dynamical process in which the $U(1)$ symmetry is restored. The
equations flow to the strong-coupling fixed point, even when starting
with symmetry broken initial conditions. At low-temperatures and
low-bias voltages the steady-state conductance approaches its
quantized backscattering free value in the general case. This is a crossover transition, characterized by a scale $T_K$, below which the edge electrons tend to screen the impurity spin and form a Kondo singlet, isotropic by its nature.

However, the RG flow process in which the system crosses over to the
low-energy isotropic regime can be cut-off before the system reaches
the strong-coupling fixed point, either by the temperature or by the
finite bias voltage. This leaves the impurity spin only partially
screened and the system accumulates a finite correction to the quantized
conductance. We studied the interplay between the
anisotropy, temperature and bias voltage in the strongly-correlated
regime numerically by employing the TD-NRG method. For $U(1)$ symmetry broken
systems, with anisotropic exchange couplings, we found that if the
temperature or bias voltage are larger than the Kondo scale, then
there is a finite backscattering current, as the impurity is only
fractionally screened. We tracked the crossover from the weak-coupling
free-moment regime to the strong-coupling screened regime,
characterized by a restored isotropic exchange and vanishing
backscattered current. The perfect conductance of the edge is restored.

The challenging numerical analysis corroborates the analytical
understanding of the role played by the global $U(1)$ symmetry in
maintaining the conduction along the edge. Furthermore, it allowed us
to extract the way in which the backscattering vanishes, and the
perfect edge conductance is restored as we reduce the bias voltage
(holding $T \ll T_K$) or reduce the temperature (holding $eV <
T_K$). In the first case, the backscattering vanished logarithmically
while $eV>T_K$, as it served as the effective cutoff for the RG
process. In the latter case, when the temperature was reduced, the
conductance follows a power-law $G\sim (T/T_K)^\alpha$ with an
exponent of $\alpha=2$, which is characteristic of a Fermi liquid
fixed point. While such an exponent cannot characterize the
linear-conductance, as it requires a time-reversal symmetry breaking
term in the low-energy fixed point Hamiltonian, we understand it
to be a feature of the nonequilibrium finite-bias condition that
explicitly break this symmetry. We expect the term to scale with
$(eV T)^2/T_K^4$, and intend to explore this further in future studies.
The $\alpha=2$ result suggests that the anisotropy
might serve as a dominant cause for the experimental observation of
non-perfect conductance in these setups.

\begin{acknowledgments}
The authors would like to thank D. Litinski, A. Bruch, E. Sela,
M. Goldstein, C. Karrasch, B. Sbierski, F. von Oppen and P.\ W. Brouwer for useful and
enlightening discussions. YVA acknowledges funding from Deutsche
Forschungsgemeinschaft (project C02 of CRC1283 and project A01 of
CRC/TR183). F.B.A. and D.M. acknowledge support from the 
Deutsche Forschungsgemeinschaft via project AN-275/8-1. 
\end{acknowledgments}

\appendix

\section{Poor man's scaling}
\label{app:poor_mans}

Here, we analyze analyze the low-energy scaling behavior of the
Hamiltonian of Eqs.~(\ref{eq:H_0_scatteringstates}) and
(\ref{eq:H_D_scatteringstates}), for $\epsilon_d = U = 0$ and
$S=1/2$. To this end, we employ Anderson's poor man's scaling. 

The model describes free fermions that couple to the impurity spin degrees of freedom with an effective Lorentzian density of states $\rho(\epsilon) = \rho_0/[1+(\epsilon/\Gamma)^2]$. Around the weak-coupling point and for simplicity, we can replace the Lorentizan density of states with a hard-cutoff density of states of with width $2\Gamma$ and $\rho_\Gamma = \pi\rho_0/2$, and ignore all the states that are outside this box. The width of the level $\Gamma$ will now serve as the new high-energy cutoff. This can be thought of as a first step in a RG process where states which have small overlap with the impurity are being integrated out. While we know that for $U\neq 0$ the width $\Gamma$ itself is a dynamic quantity that undergoes renormalization, we are working in the limit where $U=0$ and are interested in the flow of the exchange coupling, therefore we can safely omit these high-energy modes.

The next step is to rescale the Hamiltonian and the field operators with the effective bandwidth $D\equiv \Gamma$
\begin{eqnarray}
	\frac{\mathcal{H}}{D} &=&
	\sum_{\sigma} \int_{-1}^{1}\! dx x \varphi^{\dagger}_{\sigma}(x)\varphi(x) +
	\nonumber \\ &&
	\sum_{\alpha,\beta,\lambda,\lambda '}\!
	J'_{\alpha,\beta}\int_{-1}^{1}\! dx_1 dx_2 
	\varphi^{\dagger}_{\lambda}(x_1)\varphi_{\lambda '}(x_2)
	\sigma^{\alpha}_{\lambda,\lambda '} S^{\beta},
\end{eqnarray}
where $J'_{\alpha,\beta} = J_{\alpha,\beta}/v_F$, and we have defined the dimensionless field operators
\begin{equation}
	\varphi_{\sigma}(x) = \sqrt{D} \psi_{\sigma}(xD),
\end{equation}
with $\psi_{\sigma}(\epsilon)$ the on-shell energy-field operator
\begin{equation}
	\psi_{\sigma}(\epsilon) = \sqrt{\frac{D}{2}}\sum_{k}\gamma_{\sigma,k}
								\delta(\epsilon-\sigma\epsilon_k).
\end{equation}

The next step is to divide the energy band into low-energy $|x| < 1-dl$ and high-energy $1-dl < |x| \leq 1$ modes, and integrate out the fast energy modes by perturbation theory. The leading order then gives
\begin{eqnarray}
	V_{\rm eff} &=&
	-\sum_{\{\lambda_i\},\{\alpha_i\},\{\beta_i\}}
	J_{\alpha_1,\beta_2}J_{\alpha_2,\beta_2}
	\sigma^{\alpha_1}_{\lambda_1,\lambda_2}
	\sigma^{\alpha_2}_{\lambda_3,\lambda_4}
	S^{\beta_1}S^{\beta_2}
	\times \nonumber \\ &&
	\int_{-1+dl}^{1-dl}dx_{1,<}dx_{2,<}
	\varphi^{\dagger}_{\lambda_1}(x_{1,<})
	\varphi_{\lambda_4}(x_{2,<}) \times	
	\nonumber \\ &&
	\int_{1-dl}^{1}\! dx_{1,>}dx_{2,>}
	\langle\varphi_{\lambda_2}(x_>)
	\varphi^{\dagger}_{\lambda_3}(x_>)
	\rangle
\end{eqnarray}
and its corresponding contributions from the modes in $(-1,-1+dl)$. We employ the identity
\begin{equation}
	(\vec{A} \cdot \vec{S}) (\vec{B}\cdot\vec{S}) = 
	i (\vec{A}\times \vec{B})\cdot \vec{S} + \vec{A}\cdot \vec{B}
\end{equation}
to carry out the multiplications and arrive at
\begin{eqnarray}
	V_{\rm eff} &=& 2dl
	\sum_{\{\alpha_i\},\{\beta_i\},\lambda_1\lambda_2}
	\epsilon_{\alpha_1,\alpha_2,\alpha_3}
	\epsilon_{\beta_1,\beta_2,\beta_3} \times \nonumber \\ &&
	{J'}_{\alpha_1,\beta_1}{J'}_{\alpha_2,\beta_2}
	\sigma^{\alpha_3}_{\lambda_1,\lambda_2}S^{\beta_3}
	\times \nonumber \\ &&
	\int_{-1+dl}^{1-dl}\!dx_{1,<}dx_{2,<}
	\varphi^{\dagger}_{\lambda_1}(x_{1,<})
	\varphi_{\lambda_1}(x_{2,<}),
\end{eqnarray}
where we have omitted constant terms and terms contributing to a scattering potential, which are irrelevant. The above expression can be written in a more concise form if we write the exchange couplings as vectors in the impurity spin-basis, ${\bf  J}'_{\alpha} = \sum_{\beta}{J'}_{\alpha,\beta}\hat{\beta}$. We then write the effective Hamiltonian
\begin{eqnarray}
	\frac{\mathcal{H}'}{D} &=&
	\sum_{\sigma} \int_{-1+dl}^{1-dl}\! dx x 
	\varphi^{\dagger}_{\sigma}(x)\varphi(x) +
	\nonumber \\ &&
	\sum_{\{\alpha_i\},\lambda,\lambda '}
	\left[{\bf  J}'_{\alpha_1}+2dl\epsilon_{\alpha_1,\alpha_2,\alpha_3}
	\left({\bf  J}'_{\alpha_2}\times {\bf  J}'_{\alpha_3}\right)\right]
	\cdot {\bf  S} \nonumber \\ && \times
	\int_{-1+dl}^{1-dl}
	\! dx_1 dx_2 \varphi^{\dagger}_{\lambda}(x_1)\varphi_{\lambda '}(x_2)
	\sigma^{\alpha_1}_{\lambda,\lambda '}.
\end{eqnarray}

Finally, we rescale by $dx \to (1-dl)^{1/2} dx$, and write $\mathcal{H}'$ in terms of $D' = (1-dl)D$, to have
\begin{eqnarray}
	\frac{\mathcal{H}'}{D'} &=&
	\sum_{\sigma} \int_{-1}^{1}\! dx x 
	\varphi^{\dagger}_{\sigma}(x)\varphi(x) +
	\nonumber \\ &&
	\sum_{\{\alpha_i\},\lambda,\lambda '}
	\left[{\bf  J}'_{\alpha_1}+2dl\epsilon_{\alpha_1,\alpha_2,\alpha_3}
	\left({\bf  J}'_{\alpha_2}\times {\bf  J}'_{\alpha_3}\right)\right]
	\cdot {\bf  S} \nonumber \\ && \times
	\int_{-1}^{1}
	\! dx_1 dx_2 \varphi^{\dagger}_{\lambda}(x_1)\varphi_{\lambda '}(x_2)
	\sigma^{\alpha_1}_{\lambda,\lambda '}.
\end{eqnarray}
We therefore arrive at the following renormalization group flow equations
\begin{eqnarray}
	\frac{d{\bf J}_i}{dl} &=& 2\pi \rho_\Gamma \sum_{j,k}
	\epsilon_{i,j,k}{\bf  J}_j\times{\bf J}_k,
	\label{eqs_RG_Poors}
\end{eqnarray}
with the dimensions restored, and we took the relation $\rho = (2\pi v_F)^{-1}$ for a flat band. A detailed analysis of this RG equations can be found in the appendix in Ref.~(\cite{Vinkler-Aviv_2017})

\bibliography{cond_bib}

\end{document}